\documentclass[aps,prb,reprint,showpacs,superscriptaddress]{revtex4-2}

\usepackage{graphicx}
\usepackage{physics}
\usepackage{ulem}

\usepackage{amsmath,amssymb}
\usepackage{color}
\usepackage{bm}

\newcommand{\ave}[1]{\left\langle {#1} \right\rangle}

\begin{document}

\title{Anisotropic Spin-Current Spectroscopy of Ferromagnetic Superconducting Gap Symmetries}

\author{Hiroshi Funaki}
\affiliation{%
Kavli Institute for Theoretical Sciences, University of Chinese Academy of Sciences, Beijing, 100190, China.
}%
\author{Ai Yamakage}
\affiliation{
Department of Physics, Nagoya University, Nagoya 464-8602, Japan
	}
\author{Mamoru Matsuo }
\affiliation{%
Kavli Institute for Theoretical Sciences, University of Chinese Academy of Sciences, Beijing, 100190, China.
}%

\affiliation{%
CAS Center for Excellence in Topological Quantum Computation, University of Chinese Academy of Sciences, Beijing 100190, China
}%
\affiliation{%
RIKEN Center for Emergent Matter Science (CEMS), Wako, Saitama 351-0198, Japan
}%
\affiliation{%
Advanced Science Research Center, Japan Atomic Energy Agency, Tokai, 319-1195, Japan
}%

\date{\today}

\begin{abstract}
We develop a microscopic theory of tunneling spin transport at the magnetic interface between a ferromagnetic insulator (FI) and a ferromagnetic superconductor (FSC) driven by ferromagnetic resonance. 
We show that the spin susceptibilities of the FSC can be extracted from the spin currents by tuning the easy axis of the FI, and thus the spin currents can be a probe for the symmetries of the spin-triplet Cooper pairing. 
Our results will offer a route to exploiting the synergy of magnetism and superconductivities for spin devices.
\end{abstract}

\maketitle 

\section{Introduction}
Tunneling spin current in magnetic heterostructures driven by magnetization dynamics using ferromagnetic resonance (FMR) has been studied intensively in spintronics. It is widely known as the spin pumping effect\cite{tserkovnyak2005nonlocal}, a versatile way to generate the spin current in nanohybrid systems from a ferromagnet into various conducting materials. 
Recently, the spin pumping has been recognized as a quantum probe to detect magnetic properties of thin films\cite{hanSpinCurrentProbe2020} because the generated spin current can be measured very sensitively even for nano-scale thin films\cite{qiu2016spin}.  
From a theoretical point of view, the spin current reflects the spin susceptibility of adjacent materials \cite{Ohnuma2014}. This property has a significant impact on superconducting spintronics research\cite{linderSuperconductingSpintronics2015}, where a variety of conversions between Cooper pair supercurrents and spin currents have been intensively studied including triplet Cooper pair currents\cite{johnsen2021magnon}. In particular, the tunneling spin current can be utilized as a direct probe of spin excitations in the Cooper pair symmetries of conventional SCs \cite{inoueSpinPumpingSuperconductors2017,katoMicroscopicTheorySpin2019,silaevFinitefrequencySpinSusceptibility2020,silaevLargeEnhancementSpin2020,ojajarviNonlinearSpinTorque2020,simensen2021spin,bellSpinDynamicsSuperconductorFerromagnet2008,jeonEnhancedSpinPumping2018,yaoProbeSpinDynamics2018,liPossibleEvidenceSpinTransfer2018,jeonEffectMeissnerScreening2019,jeonEnhancedSpinPumping2018,jeonAbrikosovVortexNucleation2019,golovchanskiyMagnetizationDynamicsProximityCoupled2020,zhaoExploringContributionTrapped2020} and unconventional SCs \cite{Brataas2004-qm,Carreira2021-js,ominato2022ferromagnetic,ominato2022anisotropic}. 

\begin{figure}[h]
 \includegraphics[width=.9\hsize]{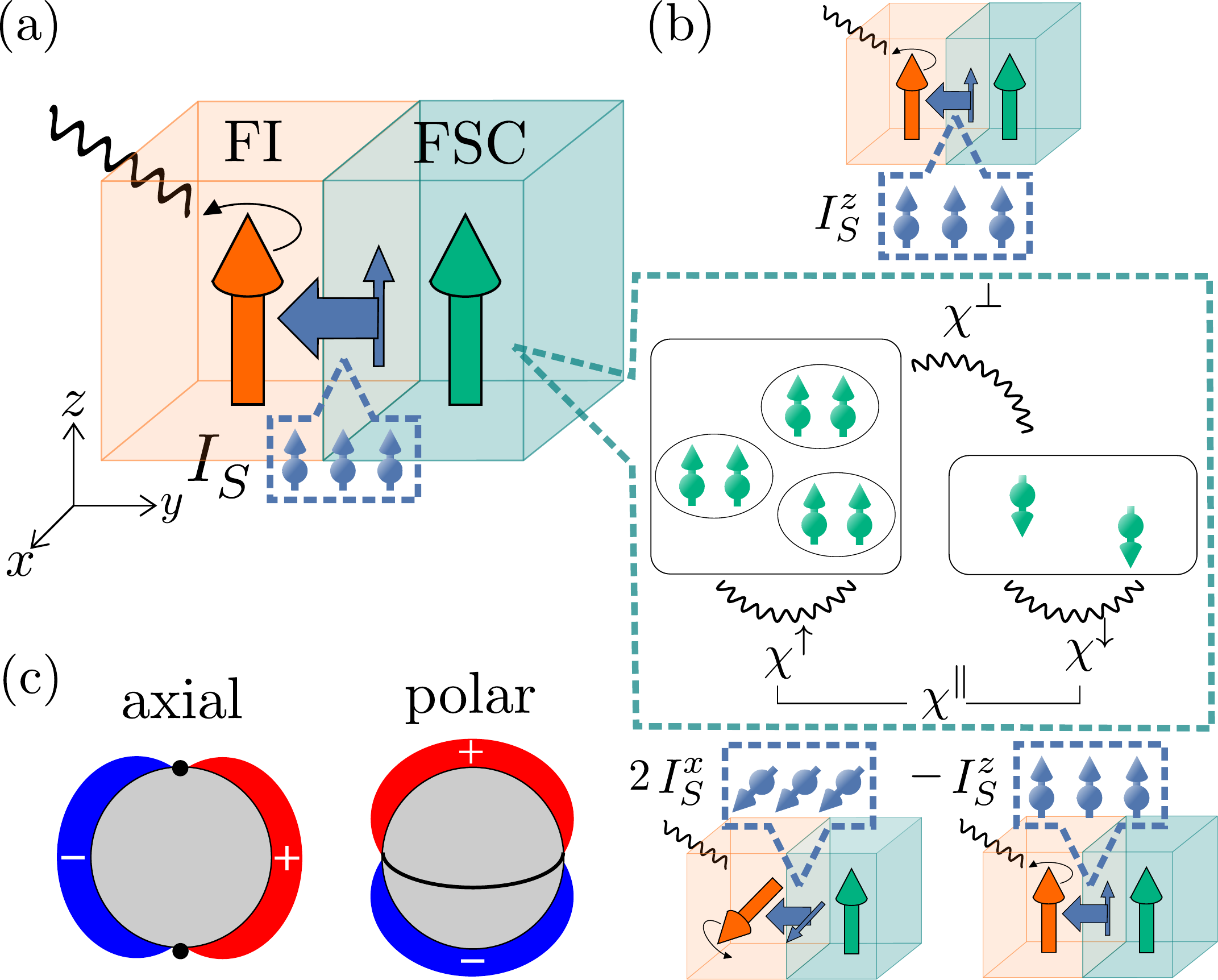}
 \caption{
 (a) Schematic diagram of the spin pumping effect at a junction system of a ferromagnetic insulator (FI) and a ferromagnetic superconductor (FSC). The tunneling spin current $I_S$ is generated at the interface driven by magnetization dynamics due to microwave irradiation in FI. (b) The spin polarization of the generated spin current can be controlled by tuning the easy axis of the magnetization in FI. The $z$-polarized spin current $I_S^z$ and the $x$-polarized one $I_S^x$ reflect the magnetic properties of the FSC characterized by the spin susceptibilities $\chi^\perp$ and $\chi^\parallel$. The transverse spin susceptibility $\chi^\perp$ can be extracted from the $z$-polarized spin current $I_S^z$ while the longitudinal one $\chi^\parallel$ from $2I_S^x-I_S^z$, where $\chi^\perp$ is the correlation between the majority spin and the minority spin and $\chi^\parallel$ consists of the spin susceptibility of the majority spins $\chi^\uparrow$ and that of the minority spins $\chi^\downarrow$.  
 (c) The superconducting gaps we consider in this paper. 
  } 
\label{schema}
\end{figure}
Symmetry of the Cooper pair characterizes the nature of superconductors \cite{Sigrist1991-vp}.
In particular, spin-triplet Cooper pairs offer a fascinating state from the viewpoint of superconducting spintronics, because they can carry spin one as a supercurrent without dissipation. 
${}^3$He is a well-established spin-triplet superfluid with (breaking) time-reversal symmetry in the B (A) phase. 
On the other hand, almost all existing superconductors are spin-singlet rather than spin-triplet superconductors. Establishing spin-triplet candidates in superconductors is an essential issue in condensed matter physics.
Indeed, the search for spin-triplet superconductors continues vigorously \cite{matano2016spin, Asaba2017-sr, Yonezawa2018-rv, Yang2021-tc, Zheng2022-mp}. 
Among superconductors, it is also believed that the spin-triplet pair is likely formed in the ferromagnetic superconducting state in uranium compounds \cite{Aoki2014-ql, Mineev2017-dw, Aoki2019-fp}, UGe$_2$ \cite{Saxena2000-xj}, UIr \cite{Akazawa2004-ku}, URhGe \cite{Aoki2001-vk}, and UCoGe \cite{Huy2007-th}.
However, despite many years of research, no definitive conclusions have been reached on its superconducting symmetry, such as the gap node and $d$-vector configuration. 
A complete characterization of the properties of ferromagnetic superconductivity will provide the basis for further understanding the physics of superconductivity and its development into anisotropic superconducting spintronics.
Moreover, it has been proposed that these ferromagnetic superconductors may exhibit gapless surface states characterized by the $\mathbb Z_4$ topological invariant under high pressure if a specific superconducting symmetry is realized \cite{daido}. 
Determining the symmetry of their parent state, ferromagnetic superconductivity, is one of the most fundamental issues for studying topological materials.

In this paper, we propose a spectroscopy of ferromagnetic superconducting gap symmetries of FSC thin films  
by using the tunnel spin current at a magnetic interface excited by FMR. 
To this end, we consider the tunneling spin transport at magnetic interface between a ferromagnetic insulator (FI) and a ferromagnetic superconductor (FSC) driven by FMR as shown in Fig.~\ref{schema}. 
We develop a microscopic theory of the spin current generation, caused by the differences of the nonequilibrium distribution functions of the FI and the FSC. We show that we can generate several types of tunneling spin current, including $I_S^z$ and $I_S^x$ by tuning the relative angle between the easy axis of the magnetization of the FI and that of the FSC. 
We find that the spin susceptibilities of the FSC can be extracted from the spin currents and propose a method to determine the symmetries of the spin-triplet Cooper pairing by using the inverse spin Hall voltage measurements. 
Our results will offer a route to exploiting the synergy of magnetism and superconductivities for spin devices.

\section {Model}Let us consider a model of tunneling spin transport at the magnetic interface consisting of a FI and a FSC aiming at extracting the magnetic properties of the FSC from the generated spin currents as shown in Fig.~\ref{schema}.  The generated spin current can be calculated by the spin tunneling Hamiltonian method \cite{Ohnuma2014,katoMicroscopicTheorySpin2019,ominato2022ferromagnetic,ominato2022anisotropic}. 
The total Hamiltonian $\mathcal{H}$ consists of the three terms:  
\begin{align}
{\cal H}=
{\cal H}_{\rm {F}SC} +{\cal H}_{\rm FI} +{\cal H}_{\rm ex}. 
\end{align}
The first term ${\cal H}_{\rm {F}SC}$ is the mean field Hamiltonian of the bulk FSC:  
\begin{align}
{\cal H}_{\rm {F}SC}
&= \frac{1}{2}\sum_{\bm k} 
  {\bm c}^\dagger_{\bm k} {\cal H}_{\rm BdG} {\bm c}_{\bm k},
\end{align}
where the fermion operator is defined by $\vb*{c}_{\vb*{k}} = (c_{\vb*{k}\uparrow}, c_{\vb*{k} \downarrow}, c_{-\vb*{k} \uparrow}^\dag, c_{-\vb*{k} \downarrow}^\dag)^{\mathrm{T}}$ in the Nambu space for $\vb*{k} = (k_x, k_y, k_z)$.
The Bogoliubov-de Gennes (BdG) Hamiltonian $\mathcal{H}_{\rm BdG}$ for a spin-triplet superconductor with an equal-spin pairing is given by 
\begin{align}
{\cal H}_{\rm BdG}=
\left(
\begin{array}{cccc}
\xi_{{\bm k} \uparrow} & 0 & \Delta_{{\bm k} \uparrow \uparrow} & 0\\
0 & \xi_{{\bm k} \downarrow} & 0 & \Delta_{{\bm k} \downarrow \downarrow}\\
-\Delta_{-{\bm k} \uparrow \uparrow}^* & 0 & -\xi_{-{\bm k} \uparrow} & 0\\
0 & -\Delta_{-{\bm k} \downarrow \downarrow}^* & 0 & -\xi_{-{\bm k} \downarrow}
\end{array}
\right), 
\end{align}
where $\xi_{{\bm k} \sigma}=\frac{\hbar^2 k^2}{2m} -\varepsilon_{\rm F} -\sigma \Delta_{\rm FM}$ is the energy dispersion of spin up [$\sigma = \uparrow (+)$] and down [$\sigma=\downarrow (-)$] electron in a ferromagnet in the normal state, and $\varepsilon_{\rm F}$ and $\Delta_{\rm FM}$ are the Fermi energy and the spin-splitting energy, respectively.
Here we focus on a single-band superconductor in order to elucidate properties intrinsic to ferromagnetic superconductivity, especially to gapless excitations of quasiparticles and nonunitarity of breaking time-reversal symmetry.
For simplicity we here neglect the collective excitations in the FSC and assume that the spin-splitting energy $\Delta_{\rm FM}$ is the constant and sufficiently larger than the superconducting gap, where only the spin-up band is superconducting, i.e., $\Delta_{{\bm k} \downarrow \downarrow}=0$.
This assumption is appropriate for examining the key characteristics of non-unitary superconductivity, a leading candidate for the order parameter in FSC.
In particular, we consider two types of the pair potential 
\begin{align}
 \Delta_{{\bm k} \uparrow \uparrow} &= -\Delta_0 \sin \theta_{\bm k} e^{-i\phi_{\bm k}}, \qfor \text{axial},
 \\
 \Delta_{{\bm k} \uparrow \uparrow} &= -\Delta_0 \cos \theta_{\bm k}, \qfor \text{polar},
\end{align}
as shown in Fig.~\ref{schema} (c).
Note that the above pair potentials are of generic form within the single-band $p$-wave equal-spin-pairing superconductivity. 
The axial gap, $A_u$ irrep of the magnetic point group $4/mm'm'$ \cite{Aroyo2011-kj, Aroyo2006-bn, Aroyo2006-jv, Elcoro2021-gs, Xu2020-lo}, has point nodes on the north and south poles while the polar gap, $E_u$ irrep, has a line node on the equator.
This difference in the excitation gap is reflected in the spin excitations involved in FMR.

The second term $\mathcal{H}_{\rm FI}$ describes the bulk FI given by 
\begin{align}
{\cal H}_{\rm FI} &=
  -J\sum_{\langle i, j \rangle} {\bm S}_i \cdot {\bm S}_j
\nonumber \\  &\quad
  -\hbar \gamma \sum_{i} \Big[ h_{\rm dc} S^z_i 
    +h_{\rm ac} ( {\rm cos} \Omega t S^x_i -{\rm sin} \Omega t S^y_i) \Big],
\end{align}
where ${\bm S}_i$ is the magnetization at the site $i$ in the FI, $J$ is the exchange interaction, $h_{\mathrm{dc}}$ is a static magnetic field, $h_{\rm ac}$ and $\Omega$ are the amplitude and frequency of the applied microwave radiation, respectively, and $\gamma$ is the gyromagnetic ratio.

The third term $\mathcal{H}_{\rm ex}$ is the interfacial exchange coupling which describes the spin transfer between the FI and the FSC: 
\begin{align}
{\cal H}_{\rm ex} &=
  { {\cal T}} \sum_{{\bm k},{\bm q}}
  \vb*{S_{{\bm k}}} \cdot \vb*{s_{{\bm q}}},
\end{align}
where ${\cal T}$ is the tunneling amplitude between the magnetization in the FI $\vb*{S_{\bm k}}$ and the conduction electron spin in the FSC $\vb*{s_{\bm q}}$.
We assumed a constant tunneling amplitude corresponding to a rough interface limit.

\section{Tunneling spin currents}
The tunneling spin current at the interface driven by FMR $\langle \hat{I}^i_S \rangle$ is calculated by the statistical average of the spin current operator $\hat{I}^i_S$ defined by
\begin{align}
\hat{I}_S^i =
  -\hbar  \partial_t (s^i_{\rm tot}) = i [s^i_{\rm tot},{\cal H}]  
 = -\sum_{{\bm k},{\bm q}} \epsilon_{i j k}
    {\cal T}  
  S_{{\bm k}}^j s_{{\bm q}}^k,   
\end{align}
where {$s^i_{\rm tot} = s^i_{{\bm q}={\bm 0}}$.
We calculate the statistical average $\langle \hat{I}^i_S \rangle$ using the Schwinger-Keldysh approach. 
By taking into account the second-order perturbation of the interfacial exchange coupling ${\cal H}_{\rm ex}$ and assuming that the Fermi energy is sufficiently larger than the spin splitting energy in FSC, we obtain the relations between the generated spin currents and the dynamic spin susceptibilities of the FSC (see the Appendices~\ref{Sec_SpC} and \ref{Sec_SpC-SSus} for the detailed derivation):
\begin{align}
I_S^{z}:=\langle \hat{I}^z_S \rangle &= 
 \frac{  {\cal T}^2}{S}
   {\rm Im}\chi_{{\rm loc}, \Omega}^{{\rm R},\perp} 
  \varDelta n_\Omega {\rm Im}G^{\rm R}_{{\bm 0},\Omega},\label{ISz}
\\
I_S^{x}:=\langle \hat{I}^x_S \rangle  &=
   \frac{{\cal T}^2}{2S}
  \qty[ {\rm Im}\chi_{{\rm loc},\Omega}^{{\rm R},\perp} 
   +{\rm Im}\chi_{{\rm loc},\Omega}^{{\rm R},\parallel} ]
  \varDelta n_\Omega {\rm Im}G^{\rm R}_{{\bm 0},\Omega},\label{ISx}
\end{align}
where 
$ \chi^{{\rm R},\perp}_{{\rm loc},\Omega}$
 and 
$ \chi^{{\rm R},\parallel}_{{\rm loc},\Omega}$ 
are transverse and longitudinal components of the local spin susceptibility in the FSC, respectively,
and
${\rm Im}G_{{\bm 0},\Omega}^{\rm R}$ 
 is spin susceptibility at ${\bm k}={\bm 0}$ in the FI, 
and $\varDelta n_{\Omega}$
 is a change of magnon number due to microwave irradiation.
Using $s^{x}$ and spin-dependent electron number $n_{\sigma}$,
$ \chi^{{\rm R},\perp}_{{\rm loc},\Omega}$
 and 
$ \chi^{{\rm R},\parallel}_{{\rm loc},\Omega}$ are described as 
$ \chi^{{\rm R},\perp}_{{\rm loc},\Omega}= 
  i \int \!\! dt \sum_{\bm q} 
  \ev{[s^{x}_{{\bm q}}(t), s^{x}_{-{\bm q}}(0)]} \theta(t) e^{i\Omega t} $
 and 
$ \chi^{{\rm R},\parallel}_{{\rm loc},\Omega}= 
  (\chi^{R,\uparrow}_{{\rm loc},\Omega} +\chi^{R,\downarrow}_{{\rm loc},\Omega})/4 $ 
 with 
$ \chi^{{\rm R},\sigma}_{{\rm loc},\Omega}= 
  i \int \!\! dt \sum_{\bm q} 
  \ev{[n_{{\bm q} \sigma}(t), n_{-{\bm q} \sigma}(0)]} \theta(t) e^{i\Omega t} $, respectively.
Equations (\ref{ISz}) and (\ref{ISx}) indicate that the local spin susceptibilities can be extracted from the spin currents as ${\rm Im}\chi^{{\rm R}, \perp}_{\rm loc} \propto I_S^z$ and ${\rm Im}\chi^{{\rm R}, \parallel}_{\rm loc} \propto 2I_S^x-I_S^z$ (see Fig.1 (b)).
The spin polarization of the spin currents can be controlled by tuning 
the relative angle between the easy axis of the magnetization of the FI and that of the FSC. 
Therefore, we can systematically identify the spin susceptibilities of the FSCs by measuring the spin currents and combining their frequency dependencies, as shown in Sec.~\ref{Sec_ome_SS}.

\begin{figure}[t]
 \includegraphics[width=1.0\hsize]{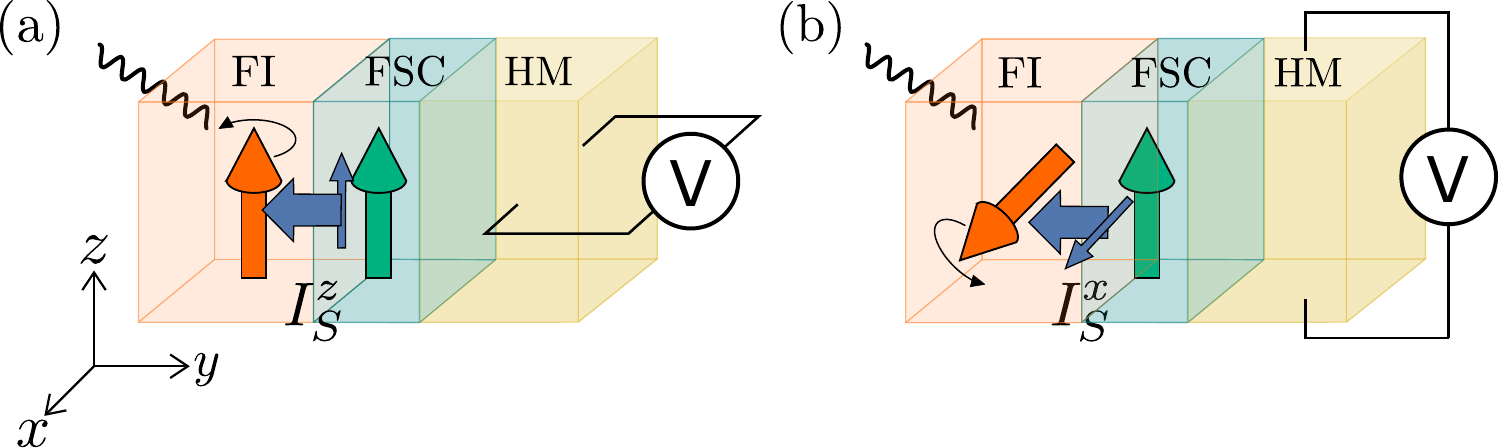}
 \caption{Schematic diagram of measuring spin currents $I_S^{z}$ (a) and $I_S^{x}$ (b) by the inverse spin Hall effect (ISHE).
 The spin current is converted to a voltage and measured.
 The conversion results from the ISHE in a heavy metal (HM) with strong spin-orbit interaction.
 The spin current at the interface between the FI and the FSC can be measured almost directly, especially when the FSC is thin.
 } 
\label{ISHEschema}
\end{figure}
We consider using the inversion spin Hall effect (ISHE) to measure the generated spin currents at the magnetic interface.
Our measurement setup is shown in Fig.~\ref{ISHEschema}, where the heavy metal (HM) with strong spin-orbit interaction (SOI), such as Pt, is attached to the FSC thin film as the spin-current detector. 
Here we assume the FSC is sufficiently thin to avoid bulk spin scattering processes in the FSC for simplicity. 
The generated spin current between the FI and the FSC flows into the HM and is converted into an inverse spin Hall voltage due to the strong SOI. 
In particular, the $z$-polarized spin current $I_S^z$ generated when the easy axis of the FI is parallel to the $z$-axis can be measured by the setup (a) in Fig.~\ref{ISHEschema}, while the $x$-polarized spin current $I_S^x$ by (b). 
Note that we can obtain sufficient information to identify the pairing symmetries of the FSC from $I_S^z$ and $I_S^x$ as discussed below. Namely, the $y$-polarized spin current, which cannot be measured in our setup, is unnecessary to determine the symmetries.

\section{Frequency dependencies of the spin susceptibilities}
\label{Sec_ome_SS}
\begin{figure*}[t]
 \includegraphics[width=1.0\hsize]{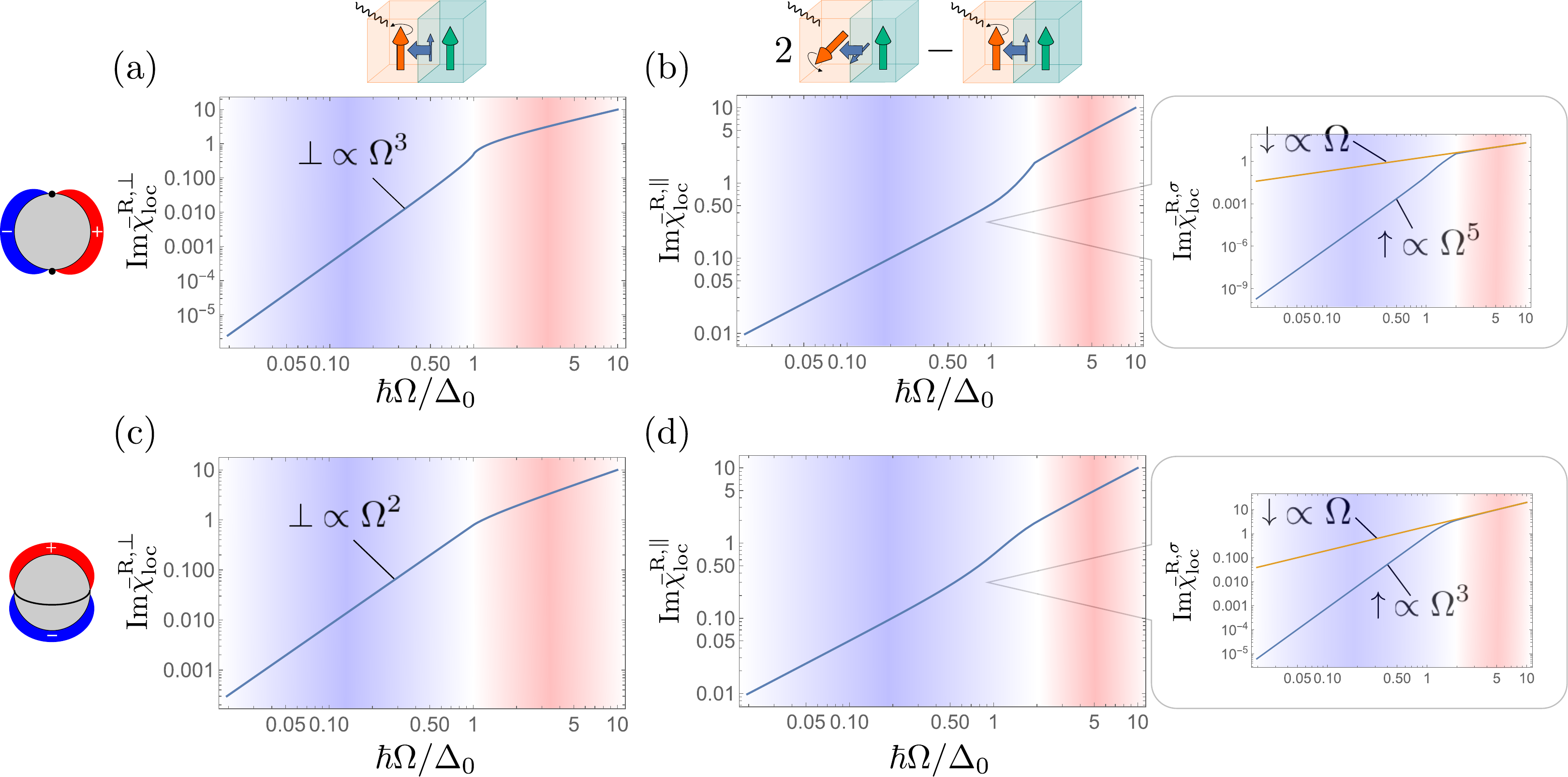}
 \caption{ The frequency dependencies of the imaginary part of the local spin susceptibilities in the axial type FSC (a,b) and those in the polar type FSC (c,d) at zero temperature. represented on a log-log scale. 
 The characteristic power-law frequency dependencies (indicated in the blue area) originate from the symmetries of the superconducting gaps. Together with the fact that ${\rm Im} \chi^{{\rm R}, \perp}_{\rm loc}$ (a,c) is extracted from $I_S^z$ and ${\rm Im} \chi^{{\rm R}, \parallel}_{\rm loc}$ (b,d) from  $2I_S^x-I_S^z$ (see also Fig. 1), we can identify the ferromagnetic superconducting gap symmetries from the tunneling spin currents.  
 }
 \label{fig_chi-ome}
\end{figure*}
%
Figure \ref{fig_chi-ome} shows the numerical results of the spin susceptibilities. 
We focus on the frequency dependence of the spin susceptibility when the temperature is lower than the frequency of the applied microwave radiation, thus the temperature can be approximated as zero.
The frequency dependencies of the imaginary part of the local spin susceptibilities in the axial type FSC ${\rm Im} \chi^{{\rm R},\perp}_{\rm loc}$(a) and ${\rm Im} \chi^{{\rm R},\parallel}_{\rm loc}$ (b) and those in the polar type FSC (c, d) at zero temperature represented on a log-log scale, 
where the spin susceptibilities are normalized as 
${\rm Im}{\bar \chi}_{\rm loc}^{{\rm R}, \alpha}={\rm Im} \chi_{{\rm loc}, \Omega}^{{\rm R}, \alpha}/{\rm Im}\chi_{{\rm loc}, {\rm N}, \Omega = \Delta_0/\hbar}^{{\rm R},\perp}$
($\alpha =\perp, \parallel, \uparrow, \downarrow$). 
Here ${\rm Im}\chi_{{\rm loc}, {\rm N}}^{{\rm R},\perp}$ is the spin susceptibility in the normal state,
and the frequency is normalized by $\Delta_0$.
The susceptibility ${\rm Im}{\bar \chi}_{\rm loc}^{{\rm R}, \alpha}$ does not depend on the spin-splitting energy $\Delta_{\rm FM}$ approximately because the Fermi energy $\varepsilon_{\rm F}$ is sufficiently larger than $\Delta_{\rm FM}$.
Thus, the normalized spin susceptibility has only one parameter, $\hbar \Omega / \Delta_0$.
It is remarkable that the spin susceptibilities show the characteristic power-law frequency dependencies. The transverse spin susceptibility in the {axial} superconducting state ${\rm Im} \chi^{{\rm R}, \perp}_{\rm loc}$ is proportional to $\Omega^3$ while that in the polar state to $\Omega^2$ in the low frequency region $\hbar \Omega \lesssim \Delta_0$, as indicated in the blue area in Figs.~2 (a) and (c).
Such frequency dependencies can be obtained from the analytical power expansions of the spin susceptibility (see the Appendix~\ref{Sec_PSE} for the detailed derivation).
For instance, the power expansion of ${\rm Im}\chi^{{\rm R}, \perp}_{\rm loc}$ in the axial superconducting state when $ \hbar \Omega \lesssim \Delta_0 $ is given by
\begin{align}
{\rm Im}{\bar \chi}_{{\rm loc},\Omega}^{R,\perp} &\approx
  \frac{(\hbar \Omega)^{3}}{3 \Delta_0^3}.
\end{align}
It should be noted that the dependence on $\Omega^3$ appears when $\hbar \Omega \lesssim \Delta_0 $, whereas it is proportional to $\Omega$ when the frequency becomes larger than the superconducting gap, where it no longer differs from the normal state.

\begin{table} 
 \caption{
The characteristic power-law frequency dependencies of the transverse (${\rm Im}\chi_{\rm loc}^{{\rm R}, \perp}$) and longitudinal 
(${\rm Im}\chi_{\rm loc}^{{\rm R}, \parallel}
=[{\rm Im}\chi^{{\rm R}, \uparrow}_{\rm loc} + {\rm Im}\chi^{{\rm R}, \downarrow}_{\rm loc}]/4$)
components of the imaginary part of the local spin susceptibilities in various states. 
The susceptibilities in the anisotropic superconducting states indicate the power-law frequency dependencies in sharp contrast to the exponential dependence in the $s$-wave spin-singlet SC. In addition to the axial and polar properties, the non-unitarity of the FSC can be identified from the frequency dependencies of the susceptibilities.   
}
 \begin{tabular}{lccc}
 \hline \hline
    SC state \quad 
  & ${\rm Im}\chi_{\rm loc}^{{\rm R}, \perp}$
  & ${\rm Im}\chi_{\rm loc}^{{\rm R}, \uparrow}$ \quad 
  & ${\rm Im}\chi_{\rm loc}^{{\rm R}, \downarrow}$ \quad \\
 \hline
    FM p-wave (axial) non-unitary & 3 & 5 & 1 \\
    FM p-wave (polar) non-unitary & 2 & 3 & 1 \\
    FM p-wave (axial) unitary     & 5 & 5 & 5 \\
    FM p-wave (polar) unitary     & 3 & 3 & 3 \\
    d-wave (2D polar) singlet     & 3 & 3 & 3 \\
    \hline
    s-wave singlet                & exp. & exp. & exp.  \\
 \hline \hline
 \end{tabular}
\label{tab_SCsymme}
\end{table}
%
In addition to the axial and polar features, the frequency dependencies in the low frequency region provide information on the non-unitarity of the FSCs. 
Due to the anisotropy of the non-unitary pair, $\Delta_{\uparrow\uparrow} \ne 0$ and $\Delta_{\downarrow\downarrow} =0$, the spin excitations are also anisotropic, making $\chi_{\rm loc}^{{\rm R}, \uparrow}$ and $\chi_{\rm loc}^{{\rm R}, \downarrow}$ different. 
Moreover, $\chi_{\rm loc}^{{\rm R}, \perp}$ differs from $\chi_{\rm loc}^{{\rm R}, \uparrow}$ and $\chi_{\rm loc}^{{\rm R}, \downarrow}$ because it depends on both spin-up and down bands. 
In particular, their exponents are also different. 
They are given by $\Im\chi_{\rm loc}^{{\rm R}, \perp} \propto \Omega^3$, $\Im\chi_{\rm loc}^{{\rm R}, \uparrow} \propto \Omega^5$, and $\Im\chi_{\rm loc}^{{\rm R}, \downarrow} \propto \Omega$ for the axial non-unitary pair. 
On the other hand, no such anisotropy is observed for the unitary pair with $\Delta_{\uparrow\uparrow}=\Delta_{\downarrow\downarrow}$, and then all components of susceptibility are proportional to the fifth power of frequency.
Thus, the measurement of spin excitations via spin currents by FMR is also helpful in measuring the non-unitary nature of the Cooper pair.

We mention two points concerning the scope of our model.
Firstly, our model does not account for the multiband of FSC. Even when the multiband is considered, the exponent of the frequency dependence remains unchanged for intraband pairings because the node structure of the superconducting gap, which are the same as for the single-band FSCs, primarily governs the exponent. When the interband paring is dominant, a different exponent is expected to be observed.
Secondly, our model does not consider the Andreev bound states because our objective is to study the spin excitation in the bulk and not the Andreev bound states emerging on the interface between the FM and FSC. They are gapless surface excitations in superconductors with gap nodes and hence can change the exponents discussed above. In the case of line nodes, Andreev bound states form in the xy-plane and do not impact the exponents. In contrast, point nodes create Andreev bound states along the $xz$ plane, which influences the exponents. In fact, a previous study has demonstrated that Andreev bound states significantly contribute to spin pumping in d-wave superconductors~\cite{Sun2022-ri}.
These issues will be left for future studies.

\section{Conclusion}
In this paper, we have developed a microscopic theory of the tunneling spin current at the magnetic interface of a FI and a FSC by microwave irradiation, aiming to identify the ferromagnetic superconducting gap symmetries. 
We obtained the relations between the tunneling spin currents and the dynamic spin susceptibilities of the FSCs, and found that the spin susceptibilities can be extracted from the spin currents by tuning the relative angle between the easy axis of the magnetization of the FI and that of the FSC. 
We revealed that the spin susceptibilities of the FSC indicate the characteristic power-law frequency dependencies reflecting the axial and polar properties as well as the non-unitarity and unitarity of the FSCs. 
Accordingly, the tunneling spin currents in our setups can be a probe of the ferromagnetic superconducting gap symmetries by combining the tunability of their spin polarization and their frequency dependencies.
Our theory paves the way for ferromagnetic superconducting spintornics, where the synergy of magnetism and superconductivities are exploited.

\acknowledgements
The authors are grateful to Yuya Ominato for valuable comments.
This work was supported by the Priority Program of the Chinese Academy of Sciences under Grant No. XDB28000000, and by JSPS KAKENHI for Grants (Nos.~20K03835, 20H04635, 20H01863, 21H01800, 21H04565, and 23H01839) from MEXT, Japan.

\appendix
\section{Spin currents at the interface of ferromagnetic junctions}
\label{Sec_SpC}
In this Section, we describe spin currents generated by spin pumping at the interface of ferromagnetic junctions.
We derive the explicit expression of the spin current of the second-order perturbation of the interfacial exchange coupling between a ferromagnetic insulator (FI) and a ferromagnetic superconductor (FSC) using the Schwinger-Keldysh approach.  

We define the spin operator in the FI, ${\bm S}_{\bm k}$, and that in the FSC, ${\bm s}_{\bm q}$, as
\begin{align}
{\bm S}_{\bm k} &= S_{\bm k}^{r} {\bm e}_{r} +S_{\bm k}^{\theta} {\bm e}_{\theta} +S_{\bm k}^{\phi} {\bm e}_{\phi},
\\
S_{\bm k}^{\pm} &= S_{\bm k}^{\theta} \pm iS_{\bm k}^{\phi},
\\
{\bm s}_{\bm q} &= s_{\bm q}^{z} {\bm e}_{z} +s_{\bm q}^{x} {\bm e}_{x} +s_{\bm q}^{y} {\bm e}_{y},
\\
s_{\bm q}^{\pm} &= s_{\bm q}^{x} \pm is_{\bm q}^{y},
\end{align}
where $S_{\bm k}^{\pm}$ and $s_{\bm q}^{\pm}$ are ladder operators, ${\bm e}_{r}=(\sin{\theta} \cos{\phi},\sin{\theta} \sin{\phi})$, ${\bm e}_{\theta}=(\cos{\theta} \cos{\phi}, \cos{\theta} \sin{\phi}, -\sin{\theta})$, ${\bm e}_{\phi}=(-\sin{\phi}, \cos{\phi}, 0)$, 
${\bm e}_{z}=(0,0,1)$, ${\bm e}_{x}=(1,0,0)$, and ${\bm e}_{y}=(0,1,0)$. Here, each quantization axis is chosen along each magnetic easy axis ${\bm e}_{r}=(\sin{\theta} \cos{\phi},\sin{\theta} \sin{\phi}, \cos{\theta})$ for the FI and ${\bm e}_{z}=(0,0,1)$ for the FSC, respectively.

The interfacial exchange coupling between the FI and the FSC is
\begin{align}
{\cal H}_{\rm ex} &=
  { {\cal T}} \sum_{{\bm k},{\bm q}}
  \vb*{S_{{\bm k}}} \cdot \vb*{s_{{\bm q}}}
 ={\cal T}  \sum_{{\bm k},{\bm q}}B_{\alpha \beta} S^{\alpha}_{\bm k} s^{\beta}_{\bm q},
\end{align}
where ${\cal T}$ is the strength of the interfacial exchange coupling and 
$B_{\alpha \beta}=
  {\bm e}_\alpha \cdot {\bm e}_\beta$. 
The spin-current operator is defined by
\begin{align}
\hat{I}_S^i &=
  -\hbar  \partial_t (s^i_{\rm tot}) = i [s^i_{\rm tot},{\cal H}]  
 = -\sum_{{\bm k},{\bm q}} \epsilon_{i j k}
    {\cal T}  
  S_{{\bm k}}^j s_{{\bm q}}^k 
\nonumber \\  &=
  -{\cal T}\sum_{{\bm k},{\bm q}}  A^{i}_{\alpha \beta} S^{\alpha}_{\bm k} s^{\beta}_{\bm q},
\end{align}
where $s^i_{\rm tot} = s^i_{{\bm q}={\bm 0}}$ and $A^{i}_{\alpha \beta} = 
  {\bm e}_i \cdot ({\bm e}_{\alpha} \times {\bm e}_{\beta}) $.

Using the Schwinger-Keldysh approach and taking into account the second-order perturbation of ${\cal T}$, the statistical average of the spin current is
\begin{align}
\langle {\hat I_S^{i}} \rangle &=
  -\frac{{\cal T}}{2} \sum_{{\bm k},{\bm q}} A^{i}_{\alpha \beta}
\langle
  T_C[s^{\beta}_{\bm q} S^{\alpha}_{\bm k}]^{\rm K}
\rangle 
\nonumber \\ &=
 -\frac{{\cal T}}{2} \sum_{{\bm k},{\bm q}} \Big[ A^{i}_{\alpha \beta}
    \langle S^{\alpha}_{\bm k} \rangle \langle s^{\beta}_{\bm q} \rangle 
\nonumber \\ & \ \ \
 +i {\cal T} \int \! \frac{d \omega}{2\pi}
  A^{i}_{\alpha \beta} B_{\alpha' \beta'}
  [\chi^{\beta \beta'}_{{\rm FSC}, {\bm q}, \omega} \chi^{\alpha \alpha'}_{{\rm FI}, {\bm k}, \omega} ]^{\rm K} \Big],
\label{SpC_1}
\end{align}
where $T_C$ is the time-ordering operator on the Keldysh contour and spin susceptibilities are defined as
\begin{align}
\chi^{\alpha \alpha'}_{{\rm FI}, {\bm k}, \omega} &=
i \int_C \! d(\tau_1-\tau_2) \big[ \langle
  T_C[S^{\alpha}_{\bm k}(\tau_1) S^{\alpha'}_{-\bm k}(\tau_2)]
\rangle 
\nonumber \\ & \ \ \ 
-i \langle S^{\alpha}_{\bm 0}(\tau_1) \rangle
   \langle S^{\alpha}_{\bm 0}(\tau_2) \rangle \big] e^{i\omega(\tau_1-\tau_2)}, 
\\
\chi^{\beta \beta'}_{{\rm FSC}, {\bm q}, \omega} &=
i \int_C \! d(\tau_1-\tau_2) \big[ \langle
  T_C[s^{\beta}_{\bm q}(\tau_1) s^{\beta'}_{-\bm q}(\tau_2)]
\rangle 
\nonumber \\ & \ \ \ 
-i \langle s^{\beta}_{\bm 0}(\tau_1) \rangle
   \langle s^{\beta}_{\bm 0}(\tau_2) \rangle \big] e^{i\omega(\tau_1-\tau_2)}, 
\end{align}
where $\int_C$ means integral on the Keldysh contour.  
The spin susceptibilities are detailed in Appendices~\ref{Sec_SSus_FI} and \ref{Sec_SSus_FSC}.

The integrand of the last term in Eq.~(\ref{SpC_1}) can be expanded with the Langreth rule as
\begin{align}
  [\chi^{\beta \beta'}_{{\rm FSC},{\bm q},\omega} \chi^{\alpha \alpha'}_{{\rm FI},{\bm k},\omega} ]^{\rm K}
&= \chi^{{\rm R}, \beta \beta'}_{{\rm FSC},{\bm q},\omega}
   \chi^{{\rm K}, \alpha \alpha'}_{{\rm FI},{\bm k},\omega}
\nonumber \\ & \ \ \ 
  +\chi^{{\rm K}, \beta \beta'}_{{\rm FSC},{\bm q},\omega}
   \chi^{{\rm A}, \alpha \alpha'}_{{\rm FI},{\bm k},\omega},
\end{align}
where each component is defined as
\begin{align}
  \chi^{{\rm K}, \beta \beta'}_{{\rm FSC},{\bm q},\omega}
&=i \int_{-\infty}^\infty \! {dt}  \big[ 
  \langle s^{\alpha}_{\bm q}(t) s^{\alpha'}_{-\bm q}(0) 
 +s^{\alpha'}_{-\bm q}(0) s^{\alpha}_{\bm q}(t) \rangle
\nonumber \\ & \ \ \ 
 -2 \langle s^{\beta}_{\bm 0}(t) \rangle  \langle s^{\beta'}_{\bm 0}(0) \rangle 
  \big] e^{i \omega t},
\\
  \chi^{{\rm R}, \beta \beta'}_{{\rm FSC},{\bm q},\omega}
&=i \int_{-\infty}^\infty \! {dt}
  \langle [s^{\alpha}_{\bm q}(t), s^{\alpha'}_{-\bm q}(0)] \rangle \theta(t)
  e^{i \omega t},
\\
  \chi^{{\rm K}, \alpha \alpha'}_{{\rm FI},{\bm k},\omega}
&=i \int_{-\infty}^\infty \! {dt}  \big[ 
  \langle S^{\alpha}_{\bm k}(t) S^{\alpha'}_{-\bm k}(0) 
 +S^{\alpha'}_{-\bm k}(0) S^{\alpha}_{\bm k}(t) \rangle 
\nonumber \\ & \ \ \ 
 -2 \langle S^{\alpha}_{\bm 0}(t) \rangle  \langle S^{\alpha'}_{\bm 0}(0) \rangle 
  \big] e^{i \omega t},
\\
  \chi^{{\rm A}, \alpha \alpha'}_{{\rm FI},{\bm k},\omega}
&=-i \int_{-\infty}^\infty \! {dt}
  \langle [S^{\alpha}_{\bm k}(t), S^{\alpha'}_{-\bm k}(0)] \rangle \theta(-t)
  e^{i \omega t}.
\end{align}
Here, the Keldysh component of the Green's function is defined by
\begin{align}
  &G^{{\rm K}}_{A,B}({\bm q},\omega) =
\nonumber \\ & \ \ \ 
i \int_{-\infty}^\infty \! {dt}  \big[ 
  \langle A_{\bm q}(t) B_{-\bm q}(0) 
 +B_{-\bm q}(0) A_{\bm q}(t) \rangle
  \big] e^{i \omega t},
\end{align}
where $A$ and $B$ are arbitrary bosonic operators.
Furthermore, the Green's function becomes pure imaginary after integrating over ${\bm k}, {\bm q}$, and $\omega$ because
$([\chi^{\beta \beta'}_{{\rm FSC},{\bm q},\omega} \chi^{\alpha \alpha'}_{{\rm FI},{\bm k},\omega} ]^{\rm K})^*
=-[\chi^{\beta \beta'}_{{\rm FSC},-{\bm q},-\omega} \chi^{\alpha \alpha'}_{{\rm FI},-{\bm k},-\omega} ]^{\rm K}$.
Accordingly, we can rewrite the spin current as
\begin{align}
\langle {\hat {\bm I}_S} \rangle &=
  -{\cal T} \sum_{{\bm k},{\bm q}}  \Big[
    ( {\bm e}_r \times {\bm e}_z )
    \langle S^{r}_{\bm k} \rangle \langle s^{z}_{\bm q} \rangle 
\nonumber \\ & \quad
  -\frac{\cal T}{2} \int \! \frac{d \omega}{2\pi}  \Big\{
    ( {\bm e}_r \times {\bm e}_z )
   \Big[ ( {\bm e}_r \cdot {\bm e}_z )
\nonumber \\ & \quad \quad \quad
   {\rm Im}[(\chi^{zz}_{{\rm FSC},{\bm q},\omega}-\chi^{xx}_{{\rm FSC},{\bm q},\omega}) (\chi^{rr}_{\rm FI}-\chi^{\theta\theta}_{{\rm FI},{\bm k},\omega})]^{\rm K}
\nonumber \\ & \quad \quad
  +{\rm Im}[\chi^{xy}_{{\rm FSC},{\bm q},\omega} \chi^{\theta\phi}_{{\rm FI},{\bm k},\omega}]^{\rm K} \Big]
\nonumber \\ & \quad \quad
   + \{ {\bm e}_r - ({\bm e}_r \cdot {\bm e}_z) {\bm e}_z \}
\nonumber \\ & \quad \quad \quad
   {\rm Im}[(\chi^{zz}_{{\rm FSC},{\bm q},\omega} +\chi^{xx}_{{\rm FSC},{\bm q},\omega}) \chi^{\theta\phi}_{{\rm FI},{\bm k},\omega}]^{\rm K}
\nonumber \\ & \quad \quad
   +2({\bm e}_r \cdot {\bm e}_z) {\bm e}_z
   {\rm Im}[\chi^{xx}_{{\rm FSC},{\bm q},\omega} \chi^{\theta\phi}_{{\rm FI},{\bm k},\omega}]^{\rm K}
\nonumber \\ & \quad \quad
   -\{ {\bm e}_z -({\bm e}_z \cdot {\bm e}_r) {\bm e}_r \}
\nonumber \\ & \quad \quad \quad
   {\rm Im}[\chi^{xy}_{{\rm FSC},{\bm q},\omega} (\chi^{rr}_{{\rm FI},{\bm k},\omega}+\chi^{\theta\theta}_{{\rm FI},{\bm k},\omega})]^{\rm K}
\nonumber \\ & \quad \quad
   -2({\bm e}_z \cdot {\bm e}_r) {\bm e}_r
   {\rm Im}[\chi^{xy}_{{\rm FSC},{\bm q},\omega} \chi^{\theta\theta}_{{\rm FI},{\bm k},\omega}]^{\rm K}
   \Big\} \Big] ,
\end{align}
where we use equations originating from the spin conservation law:
\begin{align}
\chi_{{\rm FSC},{\bm q},\omega}^{zy}&=
\chi_{{\rm FSC},{\bm q},\omega}^{yz}=
\chi_{{\rm FSC},{\bm q},\omega}^{zx}=
\chi_{{\rm FSC},{\bm q},\omega}^{xz}=0,
\\
\chi_{{\rm FSC},{\bm q},\omega}^{zz}&\neq 
\chi_{{\rm FSC},{\bm q},\omega}^{xx}=
\chi_{{\rm FSC},{\bm q},\omega}^{yy},
\\
 \chi_{{\rm FSC},{\bm q},\omega}^{xy}&=
-\chi_{{\rm FSC},{\bm q},\omega}^{yx},
\\
\chi_{{\rm FI},{\bm k},\omega}^{zy}&=
\chi_{{\rm FI},{\bm k},\omega}^{yz}=
\chi_{{\rm FI},{\bm k},\omega}^{zx}=
\chi_{{\rm FI},{\bm k},\omega}^{xz}=0,
\\
\chi_{{\rm FI},{\bm k},\omega}^{zz}&\neq 
\chi_{{\rm FI},{\bm k},\omega}^{xx}=
\chi_{{\rm FI},{\bm k},\omega}^{yy},
\\
 \chi_{{\rm FI},{\bm k},\omega}^{xy}&=
-\chi_{{\rm FI},{\bm k},\omega}^{yx},
\end{align}
In addition, we can approximate $\chi^{xy}_{\rm FSC} \approx 0$ because the Fermi energy in the FSC is usually sufficiently larger than the considered frequency.
Therefore, we obtain
\begin{align}
\langle {\hat {\bm I}_S} \rangle &\approx
  -{\cal T} \sum_{{\bm k},{\bm q}}  \Big[
    ( {\bm e}_r \times {\bm e}_z )
    \langle S^{r}_{\bm k} \rangle \langle s^{z}_{\bm q} \rangle 
\nonumber \\ & \quad
  -\frac{\cal T}{2} \int \! \frac{d \omega}{2\pi} \Big\{ 
   [ {\bm e}_r - ({\bm e}_r \cdot {\bm e}_z) {\bm e}_z ]
\nonumber \\ & \qquad \quad
   {\rm Im}[(\chi^{zz}_{{\rm FSC},{\bm q},\omega} +\chi^{xx}_{{\rm FSC},{\bm q},\omega}) \chi^{\theta\phi}_{{\rm FI},{\bm k},\omega}]^{\rm K}
\nonumber \\ & \qquad
   +2({\bm e}_r \cdot {\bm e}_z) {\bm e}_z
   {\rm Im}[\chi^{xx}_{{\rm FSC},{\bm q},\omega} \chi^{\theta\phi}_{{\rm FI},{\bm k},\omega}]^{\rm K} \Big\} \Big],
\label{eq_Is_1}
\end{align}
where we retain only the lowest order of ${\cal T}$ in each direction component of the spin current.

\section{Spin susceptibilities extracted from the measured spin currents}
\label{Sec_SpC-SSus}
In this section, we discuss how to extract the spin susceptibilities of FSC from the measured spin current.
From Eq.~(\ref{eq_Is_1}), the $z$- and $x$-polarized spin currents are 
\begin{align}
I_S^{z} &= 
  {\cal T}^2 \sum_{{\bm k},{\bm q}} \int \! \frac{d \omega}{2\pi} \Big[
   {\rm Im}{\chi^{{\rm R}, xx}_{\rm FSC}}
   {\rm Re}{\chi^{{\rm K}, \theta\phi}_{\rm FI}}
  +{\rm Im}{\chi^{{\rm K}, xx}_{\rm FSC}} 
   {\rm Re}{\chi^{{\rm A}, \theta\phi}_{\rm FI}} \Big],
\\
I_S^{x} &=
  {\cal T}^2 \sum_{{\bm k},{\bm q}} \int \! \frac{d \omega}{2\pi} \Big[
   \frac{1}{2}[{\rm Im}{\chi^{{\rm R}, xx}_{\rm FSC}} +{\rm Im}{\chi^{{\rm R}, zz}_{\rm FSC}}]
     {\rm Re}{\chi^{{\rm K}, \theta\phi}_{\rm FI}}
\nonumber \\ & \quad
  +\frac{1}{2}[{\rm Im}{\chi^{{\rm K}, xx}_{\rm FSC}} +{\rm Im}{\chi^{{\rm K}, zz}_{\rm FSC}}]
     {\rm Re}{\chi^{{\rm A}, \theta\phi}_{\rm FI}} \Big],
\end{align}
where we use that $\chi^{{\rm K}, \theta\phi}_{\rm FI}$ is a real number as shown in Appendix~\ref{Sec_SSus_FI}.
Furthermore, rewriting the spin susceptibility of the FI in terms of magnon Green's function (See Appendix~\ref{Sec_SSus_FI} for detail), we obtain
\begin{align}
I_S^{z} &= 
  {\cal T}^2 \frac{1}{S}
  [ {\rm Im}\chi_{{\rm loc}, \Omega}^{{\rm R}, \perp} ]
  \varDelta n_\Omega {\rm Im}G_{{\bm 0},\Omega}^{\rm R},
\label{eq_Is_z2}
\\
I_S^{x} &=
  {\cal T}^2 \frac{1}{2S}
  [ {\rm Im}\chi_{{\rm loc}, \Omega}^{{\rm R}, \perp} 
   +{\rm Im}\chi_{{\rm loc}, \Omega}^{{\rm R}, \parallel} ]
  \varDelta n_\Omega {\rm Im}G_{{\bm 0},\Omega}^{\rm R},
\label{eq_Is_x2}
\end{align}
where $G_{{\bm 0},\Omega}^{\rm R}$ and $\varDelta n_\Omega$ are the retarded component of magnon Green's function and the deviation in the number of magnons under the microwave radiation with frequency $\Omega$, respectively, 
and $\chi_{{\rm loc}, \Omega}^{{\rm R}, \beta\beta}=\sum_{\bm q} \chi_{{\rm FSC}, {\bm q}, \Omega}^{{\rm R}, \beta\beta}$ is the local spin susceptibility of the FSC.
The transverse $\perp$ and the longitudinal $\parallel$ components are equal to $xx$ and $zz$ components, respectively.
The spin susceptibilities of the FSC normalized by the spin susceptibility of the normal state are 
\begin{align}
\frac{{\rm Im}\chi_{{\rm loc}, \Omega}^{{\rm R}, \perp}}
     {{\rm Im}\chi_{{\rm loc},{\rm N},\Omega}^{{\rm R}, \perp} } &=
\frac{I_S^{z}}{I_{S,{\rm N}}^{z}} ,
\\
\frac{{\rm Im}\chi_{{\rm loc}, \Omega}^{{\rm R}, \parallel}}
     {{\rm Im}\chi_{{\rm loc},{\rm N},\Omega}^{{\rm R}, \perp} } &=
\frac{2I_S^{x}-I_S^{z}}{I_{S,{\rm N}}^{z}} ,
\end{align}
where
${\rm Im}\chi_{{\rm loc},{\rm N},\Omega}^{{\rm R}, \perp}$ and
$I_{S,{\rm N}}$ are the spin susceptibility and the spin current in the normal state, respectively.
The spin susceptibility of FI disappears by the normalization because it is equal in the superconducting and normal states.
These equations tell us the relationships between the spin susceptibilities of the FSC and the measured spin currents.

\section{Spin susceptibilities in ferromagnetic insulators}
\label{Sec_SSus_FI}
In this section, we derive the spin susceptibilities $\chi^{\theta \phi}_{{\rm FI}, {\bm k},\omega}$ in the FI under microwave radiation from the magnon Green's function, and show the detailed calculations for the spin currents in Eqs.~(\ref{eq_Is_z2}) and (\ref{eq_Is_x2}).

We use the Holstein-Primakoff transformation to the spin operators in the FI and employ the spin-wave approximation:
\begin{align}
S^r_{\bm k} &= (S-a_{\bm k}^\dagger a_{\bm k}),
\\
S^{+}_{\bm k} &= S^{\theta}_{\bm k} +iS^{\phi}_{\bm k} \approx \sqrt{2S} a_{\bm k},
\\
S^{-}_{\bm k} &= S^{\theta}_{\bm k} -iS^{\phi}_{\bm k} \approx \sqrt{2S} a_{\bm k}^\dagger,
\end{align}
where $a^\dagger_{\bm k}$ and $a_{\bm k}$ are the boson creation and annihilation operators, respectively.
The Hamiltonian in the FI is represented in terms of the boson operators as
\begin{align}
{\cal H}_{\rm FI} &=
  -J\sum_{\langle i, j \rangle} {\bm S}_i \cdot {\bm S}_j
\nonumber \\ & \quad
  -\hbar \gamma \sum_{i} \Big[ h_{\rm dc} S^z_i 
    +h_{\rm ac} ( {\rm cos} \Omega t S^x_i -{\rm sin} \Omega t S^y_i) \Big]
\nonumber \\  &\approx
  \sum_{\bm k} (\omega_{\bm k} +\omega_0) a^\dagger_{\bm k} a_{\bm k}
    -V_{\rm ac} ( e^{-i \Omega t}  a^\dagger_{{\bm k}={\bm 0}} +e^{i \Omega t} a_{{\bm k}={\bm 0}} ),
\end{align}
where $h_{\rm dc}$ is static magnetic field, $h_{\rm ac}$ and $\Omega$ are amplitude and frequency of applied microwave, respectively, and $J$ is the exchange coupling constant, $\langle i, j \rangle$ represents summation over all nearest-neighbor sites, $\gamma$ is the gyromagnetic ratio, and $\omega_0=\hbar \gamma h_{\rm dc}, \ V_{\rm ac}= \hbar \gamma h_{\rm ac} \sqrt{\frac{SN}{2}}$ with the number of sites $N$.
Here, we assume the parabolic magnon dispersion: $\omega_{\bm k} \propto k^2$.
The magnon Green's functions are defined as 
\begin{align}
G_{{\bm k}, \omega} &:= -i \int_C \! d(\tau_1-\tau_2)
\langle T_C[a_{\bm k}(\tau_1) a_{\bm k}^\dagger(\tau_2)] \rangle
e^{i\omega(\tau_1-\tau_2)} ,
\\
  G_{{\bm k}, \omega}^{\rm R}
&:=-i \int_{-\infty}^\infty \! {dt}
  \langle [a_{\bm k}(t), a^{\dagger}_{\bm k}(0)] \rangle 
  \theta(t) e^{i \omega t} ,
\\
  G_{{\bm k}, \omega}^{\rm K}
&:=-i \int_{-\infty}^\infty \! {dt}
  \langle a_{\bm k}(t) a^{\dagger}_{\bm k}(0) 
 +a^{\dagger}_{\bm k}(0) a_{\bm k}(t)  \rangle 
  e^{i \omega t} .
\end{align}
Introducing a phenomenological lifetime with the Gilbert damping constant $\alpha$ and considering the second-order term of $V_{\rm ac}$ as self-energy, the magnon Green's functions are written as
\begin{align}
G_{{\bm k},\omega}^{\rm R} &=
  \frac{1}{ \hbar \omega -(\omega_{\bm k} +\omega_0) + i\alpha \hbar \omega },
\\
G_{{\bm k},\omega}^{{\rm K}} &=
  G_{0,{\bm k},\omega}^{{\rm K}}
 +2 G_{{\bm k},\omega}^{\rm R}
  \Big[ -\frac{i}{\hbar} V_{\rm ac}^2 \delta_{{\bm k},{\bm 0}} 2\pi \delta(\omega -\Omega) \Big]
  G_{{\bm k},\omega}^{\rm A}
\nonumber \\ &=
  -2i {\rm Im}G_{{\bm k},\omega}^{\rm A}
  \Big[ 2n_\omega +2\delta n_{{\bm k}, \omega} +1 \Big],
\label{G_Keldysh}
\end{align}
where $G_{0}^{\rm K}$ is the non-perturbative Keldysh component, $n_\omega$
 is the Bose distribution function, and $\delta n_{{\bm k}, \omega}$ is the deviation in the Bose distribution function due to the oscillating magnetic field.
Note that the self-energy does not affect the retarded and advanced components since $V_{\rm ac}$ is a c-number.
Here, $\delta n_{{\bm k}, \omega}$ is given by
\begin{align}
\delta n_{{\bm k}, \omega}
&= \varDelta n_\omega \delta_{{\bm k},{\bm 0}} 2\pi \delta(\omega -\Omega),
\end{align}
where $\varDelta n_\omega = V_{\rm ac}^2/(2 \alpha \hbar^2 \omega)$.
The spin susceptibility $\chi_{{\rm FI}, {\bm k}, \omega}^{+-}$ is represented by
\begin{align}
\chi_{{\rm FI}, {\bm k}, \omega}^{+-}
  = -\frac{1}{2S}G_{{\bm k}, \omega},
\end{align}
where 
\begin{align}
\chi^{+ -}_{{\rm FI}, {\bm k}, \omega} := 
  i \int_C \! d(\tau_1-\tau_2)
  \langle T_C[s^+_{\bm k}(\tau_1) s^-_{-{\bm k}}(\tau_2)] \rangle
  e^{i\omega(\tau_1-\tau_2)}.
\end{align}
The $\theta \theta$, $\phi \phi$, $\theta \phi$ and $\phi \theta$ components of the spin susceptibility are rewritten as
\begin{align}
\chi_{{\rm FI}, {\bm k}, \omega}^{\theta \theta}&= 
\chi_{{\rm FI}, {\bm k}, \omega}^{\phi \phi}= 
\frac{1}{4}(\chi_{{\rm FI}, {\bm k}, \omega}^{+ -} + \chi_{{\rm FI}, {\bm k}, \omega}^{- +} ),
\\
 \chi^{\theta \phi}_{{\rm FI}, {\bm k}, \omega}&= 
-\chi^{\phi \theta}_{{\rm FI}, {\bm k}, \omega}=
\frac{i}{4}(\chi_{{\rm FI}, {\bm k}, \omega}^{+ -} - \chi_{{\rm FI}, {\bm k}, \omega}^{- +} ),
\end{align}
where 
\begin{align}
\chi^{- +}_{{\rm FI}, {\bm k}, \omega} :=
  i \int_C \! d(\tau_1-\tau_2)
  \langle T_C[s^-_{\bm k}(\tau_1) s^+_{-{\bm k}}(\tau_2) \rangle
  e^{i\omega(\tau_1-\tau_2)}.
\end{align}
The $+-$ and $-+$ components have relationships:
\begin{align}
\chi_{{\rm FI}, {\bm k}, \omega}^{{\rm R}, -+}
  &= \chi_{{\rm FI}, -{\bm k}, -\omega}^{{\rm A}, +-},
\\
\chi_{{\rm FI}, {\bm k}, \omega}^{{\rm K}, -+}
  &= \chi_{{\rm FI}, {\bm k}, \omega}^{{\rm K}, +-}.
\end{align}
Therefore, the $\theta \phi$ components of the spin susceptibility are
\begin{align}
\chi^{{\rm R}, \theta \phi}_{{\rm FI}, {\bm k},\omega} &=
 -i\frac{1}{8S} [G^{\rm R}_{{\bm k},\omega} - G^{\rm A}_{-{\bm k},-\omega}],
\label{Chi_tp_R}
\\
\chi^{{\rm K}, \theta \phi}_{{\rm FI}, {\bm k},\omega} &=
 -i\frac{1}{8S} [G^{\rm K}_{{\bm k},\omega} - G^{\rm K}_{-{\bm k},-\omega}].
\label{Chi_tp_K}
\end{align}
The Keldysh component $\chi^{{\rm K}, \theta \phi}_{{\rm FI},{\bm k},\omega}$ is real because $G^{\rm K}_{{\bm k}, \omega}$ is pure imaginary.
From Eqs.~(\ref{Chi_tp_R}), (\ref{Chi_tp_K}) and (\ref{G_Keldysh}), we obtain
\begin{align}
\chi^{{\rm K}, \theta \phi}_{{\rm FI}, {\bm k},\omega} &=
  -2(2n_\omega +1) {\rm Re} \chi^{{\rm A}, \theta \phi}_{{\rm FI}, {\bm k},\omega}
\nonumber \\ & \quad \ 
  -\frac{1}{2S}[\delta n_{{\bm k}, \omega} -\delta n_{-{\bm k}, -\omega}]
    {\rm Im}G_{{\bm 0},\Omega}^{\rm A}.
\end{align}
With this equation, we can reproduce the spin currents in Eqs.~(\ref{eq_Is_z2}) and (\ref{eq_Is_x2}).

\section{Spin susceptibilities in ferromagnetic superconductors}
\label{Sec_SSus_FSC}
In this section, we derive the spin susceptibilities in FSC, and obtain the $\uparrow$, $\downarrow$, $+-$ and $-+$ components, transform them into the $\perp$ and $\parallel$ components required to calculate the spin currents in Eqs.~(\ref{eq_Is_z2}) and (\ref{eq_Is_x2}).

The components $\sigma(= \uparrow ,\downarrow)$, $\perp$ and $\parallel$ are defined as
\begin{align}
\chi^{\sigma}_{{\rm FSC}, {\bm q},\omega}&=
i \int_C \! d(\tau_1-\tau_2)
  \langle T_C[n_{{\bm q} \sigma}(\tau_1) n_{-{\bm q} \sigma}(\tau_2)] \rangle e^{i\omega(\tau_1-\tau_2)}
\\
\chi^{\parallel}_{{\rm FSC}, {\bm q},\omega}&= \chi^{zz}_{{\rm FSC}, {\bm q},\omega},
\\
\chi^{\perp}_{{\rm FSC}, {\bm q},\omega}&= \chi^{xx}_{{\rm FSC}, {\bm q},\omega}= \chi^{yy}_{{\rm FSC}, {\bm q},\omega},
\end{align}
where 
$n_{{\bm q} \sigma}=\sum_{\bm k} c^\dagger_{{\bm k} \sigma} c_{{\bm k}+{\bm q} \sigma}$ 
and $c_{{\bm k} \sigma}$ is the annihilation operator of the electron with spin $\sigma$.

The mean field Hamiltonian in the FSC is 
\begin{align}
{\cal H}_{\rm {F}SC}
&= \frac{1}{2}\sum_{\bm k} 
  {\bm c}^\dagger_{\bm k} {\cal H}_{\rm BdG} {\bm c}_{\bm k},
\end{align}
where $\vb*{c}_{\vb*{k}} = (c_{\vb*{k}\uparrow}, c_{\vb*{k} \downarrow}, c_{-\vb*{k} \uparrow}^\dag, c_{-\vb*{k} \downarrow}^\dag)^{\mathrm{T}}$, and $\mathcal{H}_{\rm BdG}$ is 
\begin{align}
{\cal H}_{\rm BdG}=
\left(
\begin{array}{cccc}
\xi_{{\bm k} \uparrow} & 0 & \Delta_{{\bm k} \uparrow \uparrow} & 0\\
0 & \xi_{{\bm k} \downarrow} & 0 & \Delta_{{\bm k} \downarrow \downarrow}\\
-\Delta_{-{\bm k} \uparrow \uparrow}^* & 0 & -\xi_{-{\bm k} \uparrow} & 0\\
0 & -\Delta_{-{\bm k} \downarrow \downarrow}^* & 0 & -\xi_{-{\bm k} \downarrow}
\end{array}
\right), 
\end{align}
where $\xi_{{\bm k} \sigma}=\varepsilon_{\bm k} -\varepsilon_{\rm F} -\sigma \Delta_{\rm FM}$, with kinetic energy $\varepsilon_{\bm k}=\frac{\hbar^2}{2m} k^2$, the Fermi energy $\varepsilon_{\rm F}$, the spin-splitting energy $\Delta_{\rm FM}$, and $\Delta_{{\bm k} \sigma \sigma}$ is the superconducting gap.
We consider that the superconducting gap opens only in the spin-up band (i. e. $\Delta_{{\bm k} \downarrow \downarrow}=0$).
The retarded Green's functions are given by
\begin{align}
\chi^{{\rm R},\uparrow}_{{\rm FSC},{\bm q},\Omega}&= 
  i \frac{1}{N^2} \sum_{{\bm k}{\bm k}'} \int_{-\infty}^{\infty} \!\! dt \ e^{i\Omega t}
\nonumber \\ & \quad
    \ave{[c_{{\bm k} \uparrow}^\dagger c_{{\bm k}+{\bm q} \uparrow}(t),
	      c_{{\bm k}' \uparrow}^\dagger c_{{\bm k}'-{\bm q} \uparrow}(0)]}
	\theta(t)
\nonumber \\ &=
   \frac{1}{N} \sum_{\bm k}\sum_{\lambda=\pm}\sum_{\lambda'=\pm}
\nonumber \\ & \quad
  \frac{1}{4}\Big[ 1 +\frac{\xi_\uparrow \xi'_\uparrow +
    E_{\uparrow \lambda}\xi'_\uparrow +E'_{\uparrow \lambda}\xi_\uparrow +\Delta_{{\bm k} \uparrow \uparrow} \Delta_{{\bm k}+{\bm q} \uparrow \uparrow} }
    {E_{\uparrow \lambda} E'_{\uparrow \lambda'}} \Big]
\nonumber \\ & \quad
  \frac{f(E'_{\uparrow \lambda'}) -f(E_{\uparrow \lambda})}
       {\hbar \Omega + i\delta -E'_{\uparrow \lambda'} +E_{\uparrow \lambda}},
\end{align}
\begin{align}
\chi^{{\rm R},\downarrow}_{{\rm FSC},{\bm q},\Omega}&= 
  i \frac{1}{N^2} \sum_{{\bm k}{\bm k}'} \int_{-\infty}^{\infty} \!\! dt \ e^{i\Omega t}
\nonumber \\ & \quad
    \ave{[c_{{\bm k} \downarrow}^\dagger c_{{\bm k}+{\bm q} \downarrow}(t),
	      c_{{\bm k}' \downarrow}^\dagger c_{{\bm k}'-{\bm q} \downarrow}(0)]}
	\theta(t)
\nonumber \\ &=
  \frac{1}{N} \sum_{\bm k}
  \frac{f(\xi_{{\bm k}+{\bm q} \downarrow}) -f(\xi_{{\bm k} \downarrow})}{\hbar \Omega + i\delta +\varepsilon_{\bm k} -\varepsilon_{{\bm k}+{\bm q}}},
\end{align}
\begin{align}
\chi^{{\rm R}, + -}_{{\rm FSC},{\bm q},\Omega}&= 
  i \frac{1}{N^2} \sum_{{\bm k}{\bm k}'} \int_{-\infty}^{\infty} \!\! dt \ e^{i\Omega t}
\nonumber \\ & \quad
    \ave{[c_{{\bm k} \uparrow}^\dagger c_{{\bm k}+{\bm q} \downarrow}(t),
	      c_{{\bm k}' \downarrow}^\dagger c_{{\bm k}'-{\bm q} \uparrow}(0)]}
	\theta(t)
\nonumber \\ &=
  \frac{1}{N} \sum_{\bm k}\sum_{\lambda=\pm}
  \frac{1}{2}\Big( 1 +\frac{\xi'_{\uparrow}}{E'_{\uparrow \lambda}} \Big)
  \frac{f(E'_{\uparrow \lambda}) -f(\xi_{\downarrow})}
       {\hbar \Omega + i\delta -E'_{\uparrow \lambda} +\xi_{\downarrow}},
\end{align}
\begin{align}
\chi^{{\rm R}, - +}_{{\rm FSC},{\bm q},\Omega}&= 
  i \frac{1}{N^2} \sum_{{\bm k}{\bm k}'} \int_{-\infty}^{\infty} \!\! dt \ e^{i\Omega t}
\nonumber \\ & \quad
    \ave{[c_{{\bm k} \downarrow}^\dagger c_{{\bm k}+{\bm q} \uparrow}(t),
	      c_{{\bm k}' \uparrow}^\dagger c_{{\bm k}'-{\bm q} \downarrow}(0)]}
	\theta(t)
\nonumber \\ &=
  \frac{1}{N} \sum_{\bm k}\sum_{\lambda=\pm}
  \frac{1}{2}\Big( 1 +\frac{\xi_{\uparrow}}{E_{\uparrow \lambda}} \Big)
  \frac{f(\xi'_{\downarrow}) -f(E_{\uparrow \lambda})}
       {\hbar \Omega + i\delta -\xi'_{\downarrow} +E_{\uparrow \lambda}}
\nonumber \\ &=
  [\chi^{{\rm R},+-}_{{\rm FSC},-{\bm q},-\Omega}]^*,
\end{align}
where $\xi = \xi_{{\bm k} \sigma}$,
$\xi' = \xi_{{\bm k}+{\bm q} \sigma}$,
$E_{\uparrow \lambda} = \lambda \sqrt{\xi_{{\bm k} \uparrow}^2 + \Delta_{{\bm k} \uparrow \uparrow}^2}$, and
$E'_{\uparrow \lambda} = \lambda \sqrt{\xi_{{\bm k}+{\bm q} \uparrow}^2 + \Delta_{{\bm k} \uparrow \uparrow}^2}$, respectively.
To calculate the spin currents in Eqs.~(\ref{eq_Is_z2}) and (\ref{eq_Is_x2}), we only need the imaginary part of the local spin susceptibilities.
Integrating the wave number ${\bm k}$, ${\bm q}$ with 
the axial $\Delta_{{\bm k} \uparrow \uparrow} = -\Delta_0 \sin \theta_{\bm k} e^{-i\phi_{\bm k}}$ 
or
the polar $\Delta_{{\bm k} \uparrow \uparrow} = -\Delta_0 \cos \theta_{\bm k}$ type superconducting gap and using $\frac{1}{x+i\delta}=-i\pi\delta(x)+{\rm P}\frac{1}{x}$,
the imaginary part of the local spin susceptibilities are
\begin{align}
{\rm Im}\chi_{{\rm loc},\Omega}^{{\rm R}, \uparrow} &\approx 
  -\pi \int_{-\infty}^\infty \!\! d \varepsilon 
  D_{S, \varepsilon} D_{S, \varepsilon +\hbar \Omega}
  [f_{\varepsilon +\hbar \Omega} -f_{\varepsilon}], 
\\ 
{\rm Im}\chi_{{\rm loc},\Omega}^{{\rm R}, \downarrow}
 &\approx -\pi \int_{-\infty}^\infty \!\! d\varepsilon 
  {D_{{\rm F}}^\downarrow} {D_{{\rm F}}^\downarrow}
    [f_{\varepsilon +\hbar \Omega} -f_{\varepsilon}] 
\nonumber \\  &\approx
  \pi {D_{{\rm F}}^\downarrow}^2 \hbar \Omega,
\\ 
{\rm Im}\chi_{{\rm loc},\Omega}^{{\rm R}, + -} & \approx
  -\pi \int_{-\infty}^\infty \!\! d\varepsilon
  D_{{\rm F}}^\downarrow D_{S, \varepsilon +\hbar \Omega}
  [f_{\varepsilon +\hbar \Omega} -f_{\varepsilon}], 
\\ 
{\rm Im}\chi_{{\rm loc},\Omega}^{{\rm R}, - +} & \approx
  \pi \int_{-\infty}^\infty \!\! d\varepsilon
  D_{{\rm F}}^\downarrow D_{S, \varepsilon -\hbar \Omega}
  [f_{\varepsilon -\hbar \Omega} -f_{\varepsilon}]
\nonumber \\ &= 
-{\rm Im}\chi_{{\rm loc}, -\Omega}^{{\rm R} + -} = 
 {\rm Im}\chi_{{\rm loc}, \Omega}^{{\rm R} + -}.
\end{align}
Here, we replaced the wave number summation with the energy integral as 
\begin{align}
\frac{1}{N} \sum_{\bm k} (\cdots)
&\rightarrow 
  \int_{-\infty}^\infty \!\! d\varepsilon D_{\rm F}^\downarrow (\cdots),
\end{align}
for the normal state and 
\begin{align}
\frac{1}{N} \sum_{\bm k} (\cdots)
&\rightarrow \frac{D_{\rm F}^\uparrow}{4\pi} 
  \int_{-\infty}^\infty \!\! dE
  \int_{0}^\pi \!\! d\theta 
  \int_{0}^{2\pi} \!\! d\phi 
\nonumber \\ & \quad \ 
  \frac{|E| \sin{\theta}}{\sqrt{E^2 -\Delta^2(\theta, \phi)}} (\cdots)
\nonumber \\
&= \int_{-\infty}^\infty \!\! d\varepsilon D_{S,\varepsilon} (\cdots),
\end{align}
for the superconducting state, where $D_{\rm F}^\sigma$ is the spin-dependent density of states in the normal state at the Fermi energy. 

The $\perp$ component of the spin susceptibility is rewritten as
\begin{align}
\chi^{\perp}_{{\rm FSC}, {\bm q},\omega}= 
\frac{1}{4}( \chi^{+-}_{{\rm FSC}, {\bm q},\omega} + \chi^{-+}_{{\rm FSC}, {\bm q},\omega} ),
\end{align}
In itinerant electron systems, the $\parallel$ component of the spin susceptibility can be rewritten as
\begin{align}
\chi^{\parallel}_{{\rm FSC}, {\bm q},\omega}= 
\frac{1}{4}( \chi^{\uparrow}_{{\rm FSC}, {\bm q},\omega} + \chi^{\downarrow}_{{\rm FSC}, {\bm q},\omega} ).
\end{align}
Therefore, we obtain
\begin{align}
{\rm Im}\chi_{{\rm loc}, \Omega}^{{\rm R}, \parallel}&=
  \frac{\pi}{4} \hbar \Omega {D_{\rm F}^\downarrow}^2
 -\frac{\pi}{4} {D_{\rm F}^\uparrow}^2
\nonumber \\ & \quad \ \times
  \int_{-\infty}^\infty \!\! d\varepsilon
  {\bar D}_{S, \varepsilon}
  {\bar D}_{S, \varepsilon +\hbar \Omega}
  [f_{\varepsilon +\hbar \Omega} -f_{\varepsilon}],
\\
{\rm Im}\chi_{{\rm loc}, \Omega}^{{\rm R}, \perp}&=
  -\frac{\pi}{2} D_{\rm F}^\uparrow D_{\rm F}^\downarrow 
  \int_{-\infty}^\infty \!\! d\varepsilon \ 
  {\bar D}_{S, \varepsilon +\hbar \Omega}
  [f_{\varepsilon +\hbar \Omega} -f_{\varepsilon}],
\end{align}
where ${\bar D}_{S, \varepsilon}= D_{S, \varepsilon}/D_{\rm F}^\uparrow $. 
Thus, we can calculate the spin currents in Eqs.~(\ref{eq_Is_z2}) and (\ref{eq_Is_x2}) with these longitudinal and transverse components of the local spin susceptibility.

\section{Power series expansion for spin susceptibilities of FSC }
\label{Sec_PSE}
In this section, we show the power expansion of the spin susceptibility of the FSC.
The spin susceptibilities of the superconducting state are normalized by those of the normal state with the frequency fixed to be equal to $\Delta_0$ as
\begin{align}
{\rm Im}{\bar \chi}_{{\rm loc}, \Omega}^{{\rm R}, \perp}&=
\frac{{\rm Im}\chi_{{\rm loc}, \Omega}^{{\rm R}, \perp}}
     {{\rm Im}\chi_{{\rm loc}, \Omega=\Delta_0/\hbar, T > T_{\rm c}}^{{\rm R}, \perp} }
\nonumber \\ &=
  -\frac{1}{\Delta_0}
  \int_{-\infty}^\infty \!\! d\varepsilon \ 
  {\bar D}_{S, \varepsilon +\hbar \Omega}
  [ f_{\varepsilon +\hbar \Omega} -f_\varepsilon ],
\\
{\rm Im}{\bar \chi}_{{\rm loc}, \Omega}^{{\rm R}, \parallel}&=
  \frac{1}{2} \Big( \frac{1}{2} {\rm Im}{\bar \chi}_{{\rm loc}, \Omega}^{{\rm R}, \uparrow} +\frac{1}{2} {\rm Im}{\bar \chi}_{{\rm loc}, \Omega}^{{\rm R}, \downarrow} \Big),
\end{align}
\begin{align}
\frac{1}{2} {\rm Im}{\bar \chi}_{{\rm loc}, \Omega}^{{\rm R}, \uparrow} &=
\frac{1}{2} \frac{{\rm Im}\chi_{{\rm loc}, \Omega}^{{\rm R}, \uparrow}}
     {{\rm Im}\chi_{{\rm loc}, \Omega=\Delta_0/\hbar, T > T_{\rm c}}^{{\rm R}, \perp} }
\nonumber \\ &\approx
 -\frac{1}{\Delta_0} 
  \int_{-\infty}^\infty \!\! d\varepsilon \ 
  {\bar D}_{S, \varepsilon}
  {\bar D}_{S, \varepsilon +\hbar \Omega}
  [ f_{\varepsilon +\hbar \Omega} -f_\varepsilon ],
\\
\frac{1}{2} {\rm Im}{\bar \chi}_{{\rm loc}, \Omega}^{{\rm R}, \downarrow} &=
\frac{1}{2} \frac{{\rm Im}\chi_{{\rm loc}, \Omega}^{{\rm R}, \downarrow}}
     {{\rm Im}\chi_{{\rm loc}, \Omega=\Delta_0/\hbar, T > T_{\rm c}}^{{\rm R}, \perp} }
\approx
  \frac{\hbar \Omega}{\Delta_0},
\end{align}
where ${\rm Im}\chi_{{\rm loc}, \Omega=\Delta_0/\hbar, T > T_{\rm c}}^{{\rm R}, \perp}=\frac{\pi}{2} \Delta_0 D_{\rm F}^\uparrow D_{\rm F}^\downarrow$
and we use the approximation 
$D_{\rm F}^\uparrow \approx D_{\rm F}^\downarrow$, 
assuming that the spin splitting energy $\Delta_{\rm FM}$ is sufficiently smaller than the Fermi energy $\varepsilon_{\rm F}$.
In the axial superconductor, which has point nodes, the density of states is
\begin{align}
 {\bar D}_{S, \varepsilon} &= \frac{\varepsilon}{2\Delta_0} 
   {\rm ln} \Big| \frac{\varepsilon +\Delta_0}{\varepsilon -\Delta_0} \Big|.
\end{align}
When $\varepsilon \ll \Delta_0$, 
it becomes 
${\bar D}_{S, \varepsilon} \approx \frac{\varepsilon^2}{\Delta_0^2}$.
In the polar superconductor, which has line nodes, the density of states is
\begin{align}
 {\bar D}_{S, \varepsilon} = 
\left\{
\begin{array}{ll}
 \frac{\varepsilon}{\Delta_0} 
   {\rm arcsin} \frac{\Delta_0}{\varepsilon} 
 & \ \big(\frac{\Delta_0}{\varepsilon} <1 \big) \\
 \frac{\pi |\varepsilon|}{2 \Delta_0} 
 & \ \big(\frac{\Delta_0}{\varepsilon} >1 \big).
\end{array}
\right.
\end{align}
When $\Omega < \Delta_0$, we can expand the spin susceptibility to powers of frequency.
In the case of axial type, they are
\begin{align}
\frac{1}{2} {\rm Im}{\bar \chi}_{{\rm loc},\Omega}^{{\rm R},\uparrow} &\approx 
  \frac{1}{30} \frac{(\hbar \Omega)^{5}}{\Delta_0^5}, 
\\
{\rm Im}{\bar \chi}_{{\rm loc},\Omega}^{{\rm R},\perp} &\approx
  \frac{1}{3} \frac{(\hbar \Omega)^{3}}{\Delta_0^3}.
\end{align}
In the case of polar type, they are
\begin{align}
\frac{1}{2}{\rm Im}{\bar \chi}_{{\rm loc},\Omega}^{{\rm R},\uparrow} &\approx 
  \frac{\pi^2}{24} \frac{(\hbar \Omega)^{3}}{\Delta_0^3},
\\
{\rm Im}{\bar \chi}_{{\rm loc},\Omega}^{{\rm R},\perp} &\approx
  \frac{\pi}{4} \frac{|\hbar \Omega|(\hbar \Omega)}{\Delta_0^2}. 
\end{align}
Here, we expand the integral required to calculate the spin susceptibility as
\begin{align} 
& \int_{-\infty}^\infty \!\! d \varepsilon 
  D_{A, \varepsilon +\hbar \Omega/2} D_{B, \varepsilon -\hbar \Omega/2}
  [f_{\varepsilon +\hbar \Omega/2} -f_{\varepsilon -\hbar \Omega/2}] 
\nonumber \\ &=
  \int_{-\infty}^\infty \!\! d \varepsilon 
  D_{A, \varepsilon +\hbar \Omega/2} D_{B, \varepsilon -\hbar \Omega/2}
  \sum_{n=0}^\infty f^{(2n+1)} \frac{\Omega^{2n+1}}{2^{2n} (2n+1)!}
\nonumber \\ &\approx
 -\sum_{n=0}^\infty 
   \Big[ \frac{\partial^{2n}}{\partial \varepsilon^{2n}}
   ( D_{A, \varepsilon +\hbar \Omega/2} D_{B, \varepsilon -\hbar \Omega/2} )
   \Big]_{\varepsilon=0}
  \frac{\Omega^{2n+1}}{2^{2n} (2n+1)!},
\end{align}
where $D_{A,B}$ is the density of states in the superconducting or normal state, and we use $f'(\varepsilon) \approx -\delta(\varepsilon)$ assuming zero temperature.
Thus, we can analytically obtain the power exponent of the spin susceptibility of superconductors.

\bibliography{ref}

\begin{thebibliography}{44}%
\makeatletter
\providecommand \@ifxundefined [1]{%
 \@ifx{#1\undefined}
}%
\providecommand \@ifnum [1]{%
 \ifnum #1\expandafter \@firstoftwo
 \else \expandafter \@secondoftwo
 \fi
}%
\providecommand \@ifx [1]{%
 \ifx #1\expandafter \@firstoftwo
 \else \expandafter \@secondoftwo
 \fi
}%
\providecommand \natexlab [1]{#1}%
\providecommand \enquote  [1]{``#1''}%
\providecommand \bibnamefont  [1]{#1}%
\providecommand \bibfnamefont [1]{#1}%
\providecommand \citenamefont [1]{#1}%
\providecommand \href@noop [0]{\@secondoftwo}%
\providecommand \href [0]{\begingroup \@sanitize@url \@href}%
\providecommand \@href[1]{\@@startlink{#1}\@@href}%
\providecommand \@@href[1]{\endgroup#1\@@endlink}%
\providecommand \@sanitize@url [0]{\catcode `\\12\catcode `\$12\catcode
  `\&12\catcode `\#12\catcode `\^12\catcode `\_12\catcode `\%12\relax}%
\providecommand \@@startlink[1]{}%
\providecommand \@@endlink[0]{}%
\providecommand \url  [0]{\begingroup\@sanitize@url \@url }%
\providecommand \@url [1]{\endgroup\@href {#1}{\urlprefix }}%
\providecommand \urlprefix  [0]{URL }%
\providecommand \Eprint [0]{\href }%
\providecommand \doibase [0]{http://dx.doi.org/}%
\providecommand \selectlanguage [0]{\@gobble}%
\providecommand \bibinfo  [0]{\@secondoftwo}%
\providecommand \bibfield  [0]{\@secondoftwo}%
\providecommand \translation [1]{[#1]}%
\providecommand \BibitemOpen [0]{}%
\providecommand \bibitemStop [0]{}%
\providecommand \bibitemNoStop [0]{.\EOS\space}%
\providecommand \EOS [0]{\spacefactor3000\relax}%
\providecommand \BibitemShut  [1]{\csname bibitem#1\endcsname}%
\let\auto@bib@innerbib\@empty
\bibitem [{\citenamefont {Tserkovnyak}\ \emph {et~al.}(2005)\citenamefont
  {Tserkovnyak}, \citenamefont {Brataas}, \citenamefont {Bauer},\ and\
  \citenamefont {Halperin}}]{tserkovnyak2005nonlocal}%
  \BibitemOpen
  \bibfield  {author} {\bibinfo {author} {\bibfnamefont {Y.}~\bibnamefont
  {Tserkovnyak}}, \bibinfo {author} {\bibfnamefont {A.}~\bibnamefont
  {Brataas}}, \bibinfo {author} {\bibfnamefont {G.~E.}\ \bibnamefont {Bauer}},
  \ and\ \bibinfo {author} {\bibfnamefont {B.~I.}\ \bibnamefont {Halperin}},\
  }\href@noop {} {\bibfield  {journal} {\bibinfo  {journal} {Reviews of Modern
  Physics}\ }\textbf {\bibinfo {volume} {77}},\ \bibinfo {pages} {1375}
  (\bibinfo {year} {2005})}\BibitemShut {NoStop}%
\bibitem [{\citenamefont {Han}\ \emph {et~al.}(2020)\citenamefont {Han},
  \citenamefont {Maekawa},\ and\ \citenamefont
  {Xie}}]{hanSpinCurrentProbe2020}%
  \BibitemOpen
  \bibfield  {author} {\bibinfo {author} {\bibfnamefont {W.}~\bibnamefont
  {Han}}, \bibinfo {author} {\bibfnamefont {S.}~\bibnamefont {Maekawa}}, \ and\
  \bibinfo {author} {\bibfnamefont {X.-C.}\ \bibnamefont {Xie}},\ }\href
  {\doibase 10.1038/s41563-019-0456-7} {\bibfield  {journal} {\bibinfo
  {journal} {Nature Materials}\ }\textbf {\bibinfo {volume} {19}},\ \bibinfo
  {pages} {139} (\bibinfo {year} {2020})}\BibitemShut {NoStop}%
\bibitem [{\citenamefont {Qiu}\ \emph {et~al.}(2016)\citenamefont {Qiu},
  \citenamefont {Li}, \citenamefont {Hou}, \citenamefont {Arenholz},
  \citenamefont {N'Diaye}, \citenamefont {Tan}, \citenamefont {Uchida},
  \citenamefont {Sato}, \citenamefont {Okamoto}, \citenamefont {Tserkovnyak}
  \emph {et~al.}}]{qiu2016spin}%
  \BibitemOpen
  \bibfield  {author} {\bibinfo {author} {\bibfnamefont {Z.}~\bibnamefont
  {Qiu}}, \bibinfo {author} {\bibfnamefont {J.}~\bibnamefont {Li}}, \bibinfo
  {author} {\bibfnamefont {D.}~\bibnamefont {Hou}}, \bibinfo {author}
  {\bibfnamefont {E.}~\bibnamefont {Arenholz}}, \bibinfo {author}
  {\bibfnamefont {A.~T.}\ \bibnamefont {N'Diaye}}, \bibinfo {author}
  {\bibfnamefont {A.}~\bibnamefont {Tan}}, \bibinfo {author} {\bibfnamefont
  {K.-i.}\ \bibnamefont {Uchida}}, \bibinfo {author} {\bibfnamefont
  {K.}~\bibnamefont {Sato}}, \bibinfo {author} {\bibfnamefont {S.}~\bibnamefont
  {Okamoto}}, \bibinfo {author} {\bibfnamefont {Y.}~\bibnamefont
  {Tserkovnyak}},  \emph {et~al.},\ }\href@noop {} {\bibfield  {journal}
  {\bibinfo  {journal} {Nature communications}\ }\textbf {\bibinfo {volume}
  {7}},\ \bibinfo {pages} {1} (\bibinfo {year} {2016})}\BibitemShut {NoStop}%
\bibitem [{\citenamefont {Ohnuma}\ \emph {et~al.}(2014)\citenamefont {Ohnuma},
  \citenamefont {Adachi}, \citenamefont {Saitoh},\ and\ \citenamefont
  {Maekawa}}]{Ohnuma2014}%
  \BibitemOpen
  \bibfield  {author} {\bibinfo {author} {\bibfnamefont {Y.}~\bibnamefont
  {Ohnuma}}, \bibinfo {author} {\bibfnamefont {H.}~\bibnamefont {Adachi}},
  \bibinfo {author} {\bibfnamefont {E.}~\bibnamefont {Saitoh}}, \ and\ \bibinfo
  {author} {\bibfnamefont {S.}~\bibnamefont {Maekawa}},\ }\href {\doibase
  10.1103/PhysRevB.89.174417} {\bibfield  {journal} {\bibinfo  {journal} {Phys.
  Rev. B}\ }\textbf {\bibinfo {volume} {89}},\ \bibinfo {pages} {174417}
  (\bibinfo {year} {2014})}\BibitemShut {NoStop}%
\bibitem [{\citenamefont {Linder}\ and\ \citenamefont
  {Robinson}(2015)}]{linderSuperconductingSpintronics2015}%
  \BibitemOpen
  \bibfield  {author} {\bibinfo {author} {\bibfnamefont {J.}~\bibnamefont
  {Linder}}\ and\ \bibinfo {author} {\bibfnamefont {J.~W.~A.}\ \bibnamefont
  {Robinson}},\ }\href {\doibase 10.1038/nphys3242} {\bibfield  {journal}
  {\bibinfo  {journal} {Nature Physics}\ }\textbf {\bibinfo {volume} {11}},\
  \bibinfo {pages} {307} (\bibinfo {year} {2015})}\BibitemShut {NoStop}%
\bibitem [{\citenamefont {Johnsen}\ \emph {et~al.}(2021)\citenamefont
  {Johnsen}, \citenamefont {Simensen}, \citenamefont {Brataas},\ and\
  \citenamefont {Linder}}]{johnsen2021magnon}%
  \BibitemOpen
  \bibfield  {author} {\bibinfo {author} {\bibfnamefont {L.~G.}\ \bibnamefont
  {Johnsen}}, \bibinfo {author} {\bibfnamefont {H.~T.}\ \bibnamefont
  {Simensen}}, \bibinfo {author} {\bibfnamefont {A.}~\bibnamefont {Brataas}}, \
  and\ \bibinfo {author} {\bibfnamefont {J.}~\bibnamefont {Linder}},\
  }\href@noop {} {\bibfield  {journal} {\bibinfo  {journal} {Physical Review
  Letters}\ }\textbf {\bibinfo {volume} {127}},\ \bibinfo {pages} {207001}
  (\bibinfo {year} {2021})}\BibitemShut {NoStop}%
\bibitem [{\citenamefont {Inoue}\ \emph {et~al.}(2017)\citenamefont {Inoue},
  \citenamefont {Ichioka},\ and\ \citenamefont
  {Adachi}}]{inoueSpinPumpingSuperconductors2017}%
  \BibitemOpen
  \bibfield  {author} {\bibinfo {author} {\bibfnamefont {M.}~\bibnamefont
  {Inoue}}, \bibinfo {author} {\bibfnamefont {M.}~\bibnamefont {Ichioka}}, \
  and\ \bibinfo {author} {\bibfnamefont {H.}~\bibnamefont {Adachi}},\ }\href
  {\doibase 10.1103/PhysRevB.96.024414} {\bibfield  {journal} {\bibinfo
  {journal} {Physical Review B}\ }\textbf {\bibinfo {volume} {96}},\ \bibinfo
  {pages} {024414} (\bibinfo {year} {2017})}\BibitemShut {NoStop}%
\bibitem [{\citenamefont {Kato}\ \emph {et~al.}(2019)\citenamefont {Kato},
  \citenamefont {Ohnuma}, \citenamefont {Matsuo}, \citenamefont {Rech},
  \citenamefont {Jonckheere},\ and\ \citenamefont
  {Martin}}]{katoMicroscopicTheorySpin2019}%
  \BibitemOpen
  \bibfield  {author} {\bibinfo {author} {\bibfnamefont {T.}~\bibnamefont
  {Kato}}, \bibinfo {author} {\bibfnamefont {Y.}~\bibnamefont {Ohnuma}},
  \bibinfo {author} {\bibfnamefont {M.}~\bibnamefont {Matsuo}}, \bibinfo
  {author} {\bibfnamefont {J.}~\bibnamefont {Rech}}, \bibinfo {author}
  {\bibfnamefont {T.}~\bibnamefont {Jonckheere}}, \ and\ \bibinfo {author}
  {\bibfnamefont {T.}~\bibnamefont {Martin}},\ }\href {\doibase
  10.1103/PhysRevB.99.144411} {\bibfield  {journal} {\bibinfo  {journal}
  {Physical Review B}\ }\textbf {\bibinfo {volume} {99}},\ \bibinfo {pages}
  {144411} (\bibinfo {year} {2019})}\BibitemShut {NoStop}%
\bibitem [{\citenamefont
  {Silaev}(2020{\natexlab{a}})}]{silaevFinitefrequencySpinSusceptibility2020}%
  \BibitemOpen
  \bibfield  {author} {\bibinfo {author} {\bibfnamefont {M.~A.}\ \bibnamefont
  {Silaev}},\ }\href {\doibase 10.1103/PhysRevB.102.144521} {\bibfield
  {journal} {\bibinfo  {journal} {Physical Review B}\ }\textbf {\bibinfo
  {volume} {102}},\ \bibinfo {pages} {144521} (\bibinfo {year}
  {2020}{\natexlab{a}})}\BibitemShut {NoStop}%
\bibitem [{\citenamefont
  {Silaev}(2020{\natexlab{b}})}]{silaevLargeEnhancementSpin2020}%
  \BibitemOpen
  \bibfield  {author} {\bibinfo {author} {\bibfnamefont {M.~A.}\ \bibnamefont
  {Silaev}},\ }\href {\doibase 10.1103/PhysRevB.102.180502} {\bibfield
  {journal} {\bibinfo  {journal} {Physical Review B}\ }\textbf {\bibinfo
  {volume} {102}},\ \bibinfo {pages} {180502} (\bibinfo {year}
  {2020}{\natexlab{b}})}\BibitemShut {NoStop}%
\bibitem [{\citenamefont {Ojaj{\"a}rvi}\ \emph {et~al.}(2020)\citenamefont
  {Ojaj{\"a}rvi}, \citenamefont {Manninen}, \citenamefont {Heikkil{\"a}},\ and\
  \citenamefont {Virtanen}}]{ojajarviNonlinearSpinTorque2020}%
  \BibitemOpen
  \bibfield  {author} {\bibinfo {author} {\bibfnamefont {R.}~\bibnamefont
  {Ojaj{\"a}rvi}}, \bibinfo {author} {\bibfnamefont {J.}~\bibnamefont
  {Manninen}}, \bibinfo {author} {\bibfnamefont {T.~T.}\ \bibnamefont
  {Heikkil{\"a}}}, \ and\ \bibinfo {author} {\bibfnamefont {P.}~\bibnamefont
  {Virtanen}},\ }\href {\doibase 10.1103/PhysRevB.101.115406} {\bibfield
  {journal} {\bibinfo  {journal} {Physical Review B}\ }\textbf {\bibinfo
  {volume} {101}},\ \bibinfo {pages} {115406} (\bibinfo {year}
  {2020})}\BibitemShut {NoStop}%
\bibitem [{\citenamefont {Simensen}\ \emph {et~al.}(2021)\citenamefont
  {Simensen}, \citenamefont {Johnsen}, \citenamefont {Linder},\ and\
  \citenamefont {Brataas}}]{simensen2021spin}%
  \BibitemOpen
  \bibfield  {author} {\bibinfo {author} {\bibfnamefont {H.~T.}\ \bibnamefont
  {Simensen}}, \bibinfo {author} {\bibfnamefont {L.~G.}\ \bibnamefont
  {Johnsen}}, \bibinfo {author} {\bibfnamefont {J.}~\bibnamefont {Linder}}, \
  and\ \bibinfo {author} {\bibfnamefont {A.}~\bibnamefont {Brataas}},\
  }\href@noop {} {\bibfield  {journal} {\bibinfo  {journal} {Physical Review
  B}\ }\textbf {\bibinfo {volume} {103}},\ \bibinfo {pages} {024524} (\bibinfo
  {year} {2021})}\BibitemShut {NoStop}%
\bibitem [{\citenamefont {Bell}\ \emph {et~al.}(2008)\citenamefont {Bell},
  \citenamefont {Milikisyants}, \citenamefont {Huber},\ and\ \citenamefont
  {Aarts}}]{bellSpinDynamicsSuperconductorFerromagnet2008}%
  \BibitemOpen
  \bibfield  {author} {\bibinfo {author} {\bibfnamefont {C.}~\bibnamefont
  {Bell}}, \bibinfo {author} {\bibfnamefont {S.}~\bibnamefont {Milikisyants}},
  \bibinfo {author} {\bibfnamefont {M.}~\bibnamefont {Huber}}, \ and\ \bibinfo
  {author} {\bibfnamefont {J.}~\bibnamefont {Aarts}},\ }\href {\doibase
  10.1103/PhysRevLett.100.047002} {\bibfield  {journal} {\bibinfo  {journal}
  {Physical Review Letters}\ }\textbf {\bibinfo {volume} {100}},\ \bibinfo
  {pages} {047002} (\bibinfo {year} {2008})}\BibitemShut {NoStop}%
\bibitem [{\citenamefont {Jeon}\ \emph {et~al.}(2018)\citenamefont {Jeon},
  \citenamefont {Ciccarelli}, \citenamefont {Ferguson}, \citenamefont
  {Kurebayashi}, \citenamefont {Cohen}, \citenamefont {Montiel}, \citenamefont
  {Eschrig}, \citenamefont {Robinson},\ and\ \citenamefont
  {Blamire}}]{jeonEnhancedSpinPumping2018}%
  \BibitemOpen
  \bibfield  {author} {\bibinfo {author} {\bibfnamefont {K.-R.}\ \bibnamefont
  {Jeon}}, \bibinfo {author} {\bibfnamefont {C.}~\bibnamefont {Ciccarelli}},
  \bibinfo {author} {\bibfnamefont {A.~J.}\ \bibnamefont {Ferguson}}, \bibinfo
  {author} {\bibfnamefont {H.}~\bibnamefont {Kurebayashi}}, \bibinfo {author}
  {\bibfnamefont {L.~F.}\ \bibnamefont {Cohen}}, \bibinfo {author}
  {\bibfnamefont {X.}~\bibnamefont {Montiel}}, \bibinfo {author} {\bibfnamefont
  {M.}~\bibnamefont {Eschrig}}, \bibinfo {author} {\bibfnamefont {J.~W.~A.}\
  \bibnamefont {Robinson}}, \ and\ \bibinfo {author} {\bibfnamefont {M.~G.}\
  \bibnamefont {Blamire}},\ }\href {\doibase 10.1038/s41563-018-0058-9}
  {\bibfield  {journal} {\bibinfo  {journal} {Nature Materials}\ }\textbf
  {\bibinfo {volume} {17}},\ \bibinfo {pages} {499} (\bibinfo {year}
  {2018})}\BibitemShut {NoStop}%
\bibitem [{\citenamefont {Yao}\ \emph {et~al.}(2018)\citenamefont {Yao},
  \citenamefont {Song}, \citenamefont {Takamura}, \citenamefont {Cascales},
  \citenamefont {Yuan}, \citenamefont {Ma}, \citenamefont {Yun}, \citenamefont
  {Xie}, \citenamefont {Moodera},\ and\ \citenamefont
  {Han}}]{yaoProbeSpinDynamics2018}%
  \BibitemOpen
  \bibfield  {author} {\bibinfo {author} {\bibfnamefont {Y.}~\bibnamefont
  {Yao}}, \bibinfo {author} {\bibfnamefont {Q.}~\bibnamefont {Song}}, \bibinfo
  {author} {\bibfnamefont {Y.}~\bibnamefont {Takamura}}, \bibinfo {author}
  {\bibfnamefont {J.~P.}\ \bibnamefont {Cascales}}, \bibinfo {author}
  {\bibfnamefont {W.}~\bibnamefont {Yuan}}, \bibinfo {author} {\bibfnamefont
  {Y.}~\bibnamefont {Ma}}, \bibinfo {author} {\bibfnamefont {Y.}~\bibnamefont
  {Yun}}, \bibinfo {author} {\bibfnamefont {X.~C.}\ \bibnamefont {Xie}},
  \bibinfo {author} {\bibfnamefont {J.~S.}\ \bibnamefont {Moodera}}, \ and\
  \bibinfo {author} {\bibfnamefont {W.}~\bibnamefont {Han}},\ }\href {\doibase
  10.1103/PhysRevB.97.224414} {\bibfield  {journal} {\bibinfo  {journal}
  {Physical Review B}\ }\textbf {\bibinfo {volume} {97}},\ \bibinfo {pages}
  {224414} (\bibinfo {year} {2018})}\BibitemShut {NoStop}%
\bibitem [{\citenamefont {Li}\ \emph {et~al.}(2018)\citenamefont {Li},
  \citenamefont {Zhao}, \citenamefont {Zhang},\ and\ \citenamefont
  {Sun}}]{liPossibleEvidenceSpinTransfer2018}%
  \BibitemOpen
  \bibfield  {author} {\bibinfo {author} {\bibfnamefont {L.-L.}\ \bibnamefont
  {Li}}, \bibinfo {author} {\bibfnamefont {Y.-L.}\ \bibnamefont {Zhao}},
  \bibinfo {author} {\bibfnamefont {X.-X.}\ \bibnamefont {Zhang}}, \ and\
  \bibinfo {author} {\bibfnamefont {Y.}~\bibnamefont {Sun}},\ }\href {\doibase
  10.1088/0256-307X/35/7/077401} {\bibfield  {journal} {\bibinfo  {journal}
  {Chinese Physics Letters}\ }\textbf {\bibinfo {volume} {35}},\ \bibinfo
  {pages} {077401} (\bibinfo {year} {2018})}\BibitemShut {NoStop}%
\bibitem [{\citenamefont {Jeon}\ \emph
  {et~al.}(2019{\natexlab{a}})\citenamefont {Jeon}, \citenamefont {Ciccarelli},
  \citenamefont {Kurebayashi}, \citenamefont {Cohen}, \citenamefont {Montiel},
  \citenamefont {Eschrig}, \citenamefont {Wagner}, \citenamefont {Komori},
  \citenamefont {Srivastava}, \citenamefont {Robinson},\ and\ \citenamefont
  {Blamire}}]{jeonEffectMeissnerScreening2019}%
  \BibitemOpen
  \bibfield  {author} {\bibinfo {author} {\bibfnamefont {K.-R.}\ \bibnamefont
  {Jeon}}, \bibinfo {author} {\bibfnamefont {C.}~\bibnamefont {Ciccarelli}},
  \bibinfo {author} {\bibfnamefont {H.}~\bibnamefont {Kurebayashi}}, \bibinfo
  {author} {\bibfnamefont {L.~F.}\ \bibnamefont {Cohen}}, \bibinfo {author}
  {\bibfnamefont {X.}~\bibnamefont {Montiel}}, \bibinfo {author} {\bibfnamefont
  {M.}~\bibnamefont {Eschrig}}, \bibinfo {author} {\bibfnamefont
  {T.}~\bibnamefont {Wagner}}, \bibinfo {author} {\bibfnamefont
  {S.}~\bibnamefont {Komori}}, \bibinfo {author} {\bibfnamefont
  {A.}~\bibnamefont {Srivastava}}, \bibinfo {author} {\bibfnamefont {J.~W.}\
  \bibnamefont {Robinson}}, \ and\ \bibinfo {author} {\bibfnamefont {M.~G.}\
  \bibnamefont {Blamire}},\ }\href {\doibase 10.1103/PhysRevApplied.11.014061}
  {\bibfield  {journal} {\bibinfo  {journal} {Physical Review Applied}\
  }\textbf {\bibinfo {volume} {11}},\ \bibinfo {pages} {014061} (\bibinfo
  {year} {2019}{\natexlab{a}})}\BibitemShut {NoStop}%
\bibitem [{\citenamefont {Jeon}\ \emph
  {et~al.}(2019{\natexlab{b}})\citenamefont {Jeon}, \citenamefont {Ciccarelli},
  \citenamefont {Kurebayashi}, \citenamefont {Cohen}, \citenamefont {Komori},
  \citenamefont {Robinson},\ and\ \citenamefont
  {Blamire}}]{jeonAbrikosovVortexNucleation2019}%
  \BibitemOpen
  \bibfield  {author} {\bibinfo {author} {\bibfnamefont {K.-R.}\ \bibnamefont
  {Jeon}}, \bibinfo {author} {\bibfnamefont {C.}~\bibnamefont {Ciccarelli}},
  \bibinfo {author} {\bibfnamefont {H.}~\bibnamefont {Kurebayashi}}, \bibinfo
  {author} {\bibfnamefont {L.~F.}\ \bibnamefont {Cohen}}, \bibinfo {author}
  {\bibfnamefont {S.}~\bibnamefont {Komori}}, \bibinfo {author} {\bibfnamefont
  {J.~W.~A.}\ \bibnamefont {Robinson}}, \ and\ \bibinfo {author} {\bibfnamefont
  {M.~G.}\ \bibnamefont {Blamire}},\ }\href {\doibase
  10.1103/PhysRevB.99.144503} {\bibfield  {journal} {\bibinfo  {journal}
  {Physical Review B}\ }\textbf {\bibinfo {volume} {99}},\ \bibinfo {pages}
  {144503} (\bibinfo {year} {2019}{\natexlab{b}})}\BibitemShut {NoStop}%
\bibitem [{\citenamefont {Golovchanskiy}\ \emph {et~al.}(2020)\citenamefont
  {Golovchanskiy}, \citenamefont {Abramov}, \citenamefont {Stolyarov},
  \citenamefont {Chichkov}, \citenamefont {Silaev}, \citenamefont {Shchetinin},
  \citenamefont {Golubov}, \citenamefont {Ryazanov}, \citenamefont {Ustinov},\
  and\ \citenamefont
  {Kupriyanov}}]{golovchanskiyMagnetizationDynamicsProximityCoupled2020}%
  \BibitemOpen
  \bibfield  {author} {\bibinfo {author} {\bibfnamefont {I.}~\bibnamefont
  {Golovchanskiy}}, \bibinfo {author} {\bibfnamefont {N.}~\bibnamefont
  {Abramov}}, \bibinfo {author} {\bibfnamefont {V.}~\bibnamefont {Stolyarov}},
  \bibinfo {author} {\bibfnamefont {V.}~\bibnamefont {Chichkov}}, \bibinfo
  {author} {\bibfnamefont {M.}~\bibnamefont {Silaev}}, \bibinfo {author}
  {\bibfnamefont {I.}~\bibnamefont {Shchetinin}}, \bibinfo {author}
  {\bibfnamefont {A.}~\bibnamefont {Golubov}}, \bibinfo {author} {\bibfnamefont
  {V.}~\bibnamefont {Ryazanov}}, \bibinfo {author} {\bibfnamefont
  {A.}~\bibnamefont {Ustinov}}, \ and\ \bibinfo {author} {\bibfnamefont
  {M.}~\bibnamefont {Kupriyanov}},\ }\href {\doibase
  10.1103/PhysRevApplied.14.024086} {\bibfield  {journal} {\bibinfo  {journal}
  {Physical Review Applied}\ }\textbf {\bibinfo {volume} {14}},\ \bibinfo
  {pages} {024086} (\bibinfo {year} {2020})}\BibitemShut {NoStop}%
\bibitem [{\citenamefont {Zhao}\ \emph {et~al.}(2020)\citenamefont {Zhao},
  \citenamefont {Yuan}, \citenamefont {Fan},\ and\ \citenamefont
  {Zhou}}]{zhaoExploringContributionTrapped2020}%
  \BibitemOpen
  \bibfield  {author} {\bibinfo {author} {\bibfnamefont {Y.}~\bibnamefont
  {Zhao}}, \bibinfo {author} {\bibfnamefont {Y.}~\bibnamefont {Yuan}}, \bibinfo
  {author} {\bibfnamefont {K.}~\bibnamefont {Fan}}, \ and\ \bibinfo {author}
  {\bibfnamefont {Y.}~\bibnamefont {Zhou}},\ }\href {\doibase
  10.35848/1882-0786/ab721c} {\bibfield  {journal} {\bibinfo  {journal}
  {Applied Physics Express}\ }\textbf {\bibinfo {volume} {13}},\ \bibinfo
  {pages} {033002} (\bibinfo {year} {2020})}\BibitemShut {NoStop}%
\bibitem [{\citenamefont {Brataas}\ and\ \citenamefont
  {Tserkovnyak}(2004)}]{Brataas2004-qm}%
  \BibitemOpen
  \bibfield  {author} {\bibinfo {author} {\bibfnamefont {A.}~\bibnamefont
  {Brataas}}\ and\ \bibinfo {author} {\bibfnamefont {Y.}~\bibnamefont
  {Tserkovnyak}},\ }\href {\doibase 10.1103/PhysRevLett.93.087201} {\bibfield
  {journal} {\bibinfo  {journal} {Phys. Rev. Lett.}\ }\textbf {\bibinfo
  {volume} {93}},\ \bibinfo {pages} {087201} (\bibinfo {year}
  {2004})}\BibitemShut {NoStop}%
\bibitem [{\citenamefont {Carreira}\ \emph {et~al.}(2021)\citenamefont
  {Carreira}, \citenamefont {Sanchez-Manzano}, \citenamefont {Yoo},
  \citenamefont {Seurre}, \citenamefont {Rouco}, \citenamefont {Sander},
  \citenamefont {Santamar\'{\i}a}, \citenamefont {Anane},\ and\ \citenamefont
  {Villegas}}]{Carreira2021-js}%
  \BibitemOpen
  \bibfield  {author} {\bibinfo {author} {\bibfnamefont {S.~J.}\ \bibnamefont
  {Carreira}}, \bibinfo {author} {\bibfnamefont {D.}~\bibnamefont
  {Sanchez-Manzano}}, \bibinfo {author} {\bibfnamefont {M.-W.}\ \bibnamefont
  {Yoo}}, \bibinfo {author} {\bibfnamefont {K.}~\bibnamefont {Seurre}},
  \bibinfo {author} {\bibfnamefont {V.}~\bibnamefont {Rouco}}, \bibinfo
  {author} {\bibfnamefont {A.}~\bibnamefont {Sander}}, \bibinfo {author}
  {\bibfnamefont {J.}~\bibnamefont {Santamar\'{\i}a}}, \bibinfo {author}
  {\bibfnamefont {A.}~\bibnamefont {Anane}}, \ and\ \bibinfo {author}
  {\bibfnamefont {J.~E.}\ \bibnamefont {Villegas}},\ }\href {\doibase
  10.1103/PhysRevB.104.144428} {\bibfield  {journal} {\bibinfo  {journal}
  {Phys. Rev. B}\ }\textbf {\bibinfo {volume} {104}},\ \bibinfo {pages}
  {144428} (\bibinfo {year} {2021})}\BibitemShut {NoStop}%
\bibitem [{\citenamefont {Ominato}\ \emph
  {et~al.}(2022{\natexlab{a}})\citenamefont {Ominato}, \citenamefont
  {Yamakage}, \citenamefont {Kato},\ and\ \citenamefont
  {Matsuo}}]{ominato2022ferromagnetic}%
  \BibitemOpen
  \bibfield  {author} {\bibinfo {author} {\bibfnamefont {Y.}~\bibnamefont
  {Ominato}}, \bibinfo {author} {\bibfnamefont {A.}~\bibnamefont {Yamakage}},
  \bibinfo {author} {\bibfnamefont {T.}~\bibnamefont {Kato}}, \ and\ \bibinfo
  {author} {\bibfnamefont {M.}~\bibnamefont {Matsuo}},\ }\href@noop {}
  {\bibfield  {journal} {\bibinfo  {journal} {Physical Review B}\ }\textbf
  {\bibinfo {volume} {105}},\ \bibinfo {pages} {205406} (\bibinfo {year}
  {2022}{\natexlab{a}})}\BibitemShut {NoStop}%
\bibitem [{\citenamefont {Ominato}\ \emph
  {et~al.}(2022{\natexlab{b}})\citenamefont {Ominato}, \citenamefont
  {Yamakage},\ and\ \citenamefont {Matsuo}}]{ominato2022anisotropic}%
  \BibitemOpen
  \bibfield  {author} {\bibinfo {author} {\bibfnamefont {Y.}~\bibnamefont
  {Ominato}}, \bibinfo {author} {\bibfnamefont {A.}~\bibnamefont {Yamakage}}, \
  and\ \bibinfo {author} {\bibfnamefont {M.}~\bibnamefont {Matsuo}},\
  }\href@noop {} {\bibfield  {journal} {\bibinfo  {journal} {Physical Review
  B}\ }\textbf {\bibinfo {volume} {106}},\ \bibinfo {pages} {L161406} (\bibinfo
  {year} {2022}{\natexlab{b}})}\BibitemShut {NoStop}%
\bibitem [{\citenamefont {Sigrist}\ and\ \citenamefont
  {Ueda}(1991)}]{Sigrist1991-vp}%
  \BibitemOpen
  \bibfield  {author} {\bibinfo {author} {\bibfnamefont {M.}~\bibnamefont
  {Sigrist}}\ and\ \bibinfo {author} {\bibfnamefont {K.}~\bibnamefont {Ueda}},\
  }\href {\doibase 10.1103/RevModPhys.63.239} {\bibfield  {journal} {\bibinfo
  {journal} {Rev. Mod. Phys.}\ }\textbf {\bibinfo {volume} {63}},\ \bibinfo
  {pages} {239} (\bibinfo {year} {1991})}\BibitemShut {NoStop}%
\bibitem [{\citenamefont {Matano}\ \emph {et~al.}(2016)\citenamefont {Matano},
  \citenamefont {Kriener}, \citenamefont {Segawa}, \citenamefont {Ando},\ and\
  \citenamefont {Zheng}}]{matano2016spin}%
  \BibitemOpen
  \bibfield  {author} {\bibinfo {author} {\bibfnamefont {K.}~\bibnamefont
  {Matano}}, \bibinfo {author} {\bibfnamefont {M.}~\bibnamefont {Kriener}},
  \bibinfo {author} {\bibfnamefont {K.}~\bibnamefont {Segawa}}, \bibinfo
  {author} {\bibfnamefont {Y.}~\bibnamefont {Ando}}, \ and\ \bibinfo {author}
  {\bibfnamefont {G.-q.}\ \bibnamefont {Zheng}},\ }\href@noop {} {\bibfield
  {journal} {\bibinfo  {journal} {Nat. Phys.}\ }\textbf {\bibinfo {volume}
  {12}},\ \bibinfo {pages} {852} (\bibinfo {year} {2016})}\BibitemShut
  {NoStop}%
\bibitem [{\citenamefont {Asaba}\ \emph {et~al.}(2017)\citenamefont {Asaba},
  \citenamefont {Lawson}, \citenamefont {Tinsman}, \citenamefont {Chen},
  \citenamefont {Corbae}, \citenamefont {Li}, \citenamefont {Qiu},
  \citenamefont {Hor}, \citenamefont {Fu},\ and\ \citenamefont
  {Li}}]{Asaba2017-sr}%
  \BibitemOpen
  \bibfield  {author} {\bibinfo {author} {\bibfnamefont {T.}~\bibnamefont
  {Asaba}}, \bibinfo {author} {\bibfnamefont {B.~J.}\ \bibnamefont {Lawson}},
  \bibinfo {author} {\bibfnamefont {C.}~\bibnamefont {Tinsman}}, \bibinfo
  {author} {\bibfnamefont {L.}~\bibnamefont {Chen}}, \bibinfo {author}
  {\bibfnamefont {P.}~\bibnamefont {Corbae}}, \bibinfo {author} {\bibfnamefont
  {G.}~\bibnamefont {Li}}, \bibinfo {author} {\bibfnamefont {Y.}~\bibnamefont
  {Qiu}}, \bibinfo {author} {\bibfnamefont {Y.~S.}\ \bibnamefont {Hor}},
  \bibinfo {author} {\bibfnamefont {L.}~\bibnamefont {Fu}}, \ and\ \bibinfo
  {author} {\bibfnamefont {L.}~\bibnamefont {Li}},\ }\href@noop {} {\bibfield
  {journal} {\bibinfo  {journal} {Phys. Rev. X}\ }\textbf {\bibinfo {volume}
  {7}} (\bibinfo {year} {2017})}\BibitemShut {NoStop}%
\bibitem [{\citenamefont {Yonezawa}(2018)}]{Yonezawa2018-rv}%
  \BibitemOpen
  \bibfield  {author} {\bibinfo {author} {\bibfnamefont {S.}~\bibnamefont
  {Yonezawa}},\ }\href@noop {} {\bibfield  {journal} {\bibinfo  {journal}
  {Condens. Matter}\ }\textbf {\bibinfo {volume} {4}},\ \bibinfo {pages} {2}
  (\bibinfo {year} {2018})}\BibitemShut {NoStop}%
\bibitem [{\citenamefont {Yang}\ \emph {et~al.}(2021)\citenamefont {Yang},
  \citenamefont {Luo}, \citenamefont {Yi}, \citenamefont {Shi}, \citenamefont
  {Zhou},\ and\ \citenamefont {Zheng}}]{Yang2021-tc}%
  \BibitemOpen
  \bibfield  {author} {\bibinfo {author} {\bibfnamefont {J.}~\bibnamefont
  {Yang}}, \bibinfo {author} {\bibfnamefont {J.}~\bibnamefont {Luo}}, \bibinfo
  {author} {\bibfnamefont {C.}~\bibnamefont {Yi}}, \bibinfo {author}
  {\bibfnamefont {Y.}~\bibnamefont {Shi}}, \bibinfo {author} {\bibfnamefont
  {Y.}~\bibnamefont {Zhou}}, \ and\ \bibinfo {author} {\bibfnamefont {G.-q.}\
  \bibnamefont {Zheng}},\ }\href@noop {} {\bibfield  {journal} {\bibinfo
  {journal} {Sci. Adv.}\ }\textbf {\bibinfo {volume} {7}},\ \bibinfo {pages}
  {eabl4432} (\bibinfo {year} {2021})}\BibitemShut {NoStop}%
\bibitem [{\citenamefont {Zheng}()}]{Zheng2022-mp}%
  \BibitemOpen
  \bibfield  {author} {\bibinfo {author} {\bibfnamefont {G.-q.}\ \bibnamefont
  {Zheng}},\ }\href@noop {} {\enquote {\bibinfo {title} {{High temperature
  spin-triplet topological superconductivity in K$_2$Cr$_3$As$_3$}},}\ }\Eprint
  {http://arxiv.org/abs/2212.02074} {arXiv:2212.02074} \BibitemShut {NoStop}%
\bibitem [{\citenamefont {Aoki}\ and\ \citenamefont
  {Flouquet}(2014)}]{Aoki2014-ql}%
  \BibitemOpen
  \bibfield  {author} {\bibinfo {author} {\bibfnamefont {D.}~\bibnamefont
  {Aoki}}\ and\ \bibinfo {author} {\bibfnamefont {J.}~\bibnamefont
  {Flouquet}},\ }\href {\doibase 10.7566/JPSJ.83.061011} {\bibfield  {journal}
  {\bibinfo  {journal} {J. Phys. Soc. Jpn.}\ }\textbf {\bibinfo {volume}
  {83}},\ \bibinfo {pages} {061011} (\bibinfo {year} {2014})}\BibitemShut
  {NoStop}%
\bibitem [{\citenamefont {Mineev}(2017)}]{Mineev2017-dw}%
  \BibitemOpen
  \bibfield  {author} {\bibinfo {author} {\bibfnamefont {V.~P.}\ \bibnamefont
  {Mineev}},\ }\href {\doibase 10.3367/UFNe.2016.04.037771} {\bibfield
  {journal} {\bibinfo  {journal} {Phys. Usp.}\ }\textbf {\bibinfo {volume}
  {60}},\ \bibinfo {pages} {121} (\bibinfo {year} {2017})}\BibitemShut
  {NoStop}%
\bibitem [{\citenamefont {Aoki}\ \emph {et~al.}(2019)\citenamefont {Aoki},
  \citenamefont {Ishida},\ and\ \citenamefont {Flouquet}}]{Aoki2019-fp}%
  \BibitemOpen
  \bibfield  {author} {\bibinfo {author} {\bibfnamefont {D.}~\bibnamefont
  {Aoki}}, \bibinfo {author} {\bibfnamefont {K.}~\bibnamefont {Ishida}}, \ and\
  \bibinfo {author} {\bibfnamefont {J.}~\bibnamefont {Flouquet}},\ }\href
  {\doibase 10.7566/JPSJ.88.022001} {\bibfield  {journal} {\bibinfo  {journal}
  {J. Phys. Soc. Jpn.}\ }\textbf {\bibinfo {volume} {88}},\ \bibinfo {pages}
  {022001} (\bibinfo {year} {2019})}\BibitemShut {NoStop}%
\bibitem [{\citenamefont {Saxena}\ \emph {et~al.}(2000)\citenamefont {Saxena},
  \citenamefont {Agarwal}, \citenamefont {Ahilan}, \citenamefont {Grosche},
  \citenamefont {Haselwimmer}, \citenamefont {Steiner}, \citenamefont {Pugh},
  \citenamefont {Walker}, \citenamefont {Julian}, \citenamefont {Monthoux},
  \citenamefont {Lonzarich}, \citenamefont {Huxley}, \citenamefont {Sheikin},
  \citenamefont {Braithwaite},\ and\ \citenamefont {Flouquet}}]{Saxena2000-xj}%
  \BibitemOpen
  \bibfield  {author} {\bibinfo {author} {\bibfnamefont {S.~S.}\ \bibnamefont
  {Saxena}}, \bibinfo {author} {\bibfnamefont {P.}~\bibnamefont {Agarwal}},
  \bibinfo {author} {\bibfnamefont {K.}~\bibnamefont {Ahilan}}, \bibinfo
  {author} {\bibfnamefont {F.~M.}\ \bibnamefont {Grosche}}, \bibinfo {author}
  {\bibfnamefont {R.~K.}\ \bibnamefont {Haselwimmer}}, \bibinfo {author}
  {\bibfnamefont {M.~J.}\ \bibnamefont {Steiner}}, \bibinfo {author}
  {\bibfnamefont {E.}~\bibnamefont {Pugh}}, \bibinfo {author} {\bibfnamefont
  {I.~R.}\ \bibnamefont {Walker}}, \bibinfo {author} {\bibfnamefont {S.~R.}\
  \bibnamefont {Julian}}, \bibinfo {author} {\bibfnamefont {P.}~\bibnamefont
  {Monthoux}}, \bibinfo {author} {\bibfnamefont {G.~G.}\ \bibnamefont
  {Lonzarich}}, \bibinfo {author} {\bibfnamefont {A.}~\bibnamefont {Huxley}},
  \bibinfo {author} {\bibfnamefont {I.}~\bibnamefont {Sheikin}, \bibfnamefont
  {I}}, \bibinfo {author} {\bibfnamefont {D.}~\bibnamefont {Braithwaite}}, \
  and\ \bibinfo {author} {\bibfnamefont {J.}~\bibnamefont {Flouquet}},\ }\href
  {\doibase 10.1038/35020500} {\bibfield  {journal} {\bibinfo  {journal}
  {Nature}\ }\textbf {\bibinfo {volume} {406}},\ \bibinfo {pages} {587}
  (\bibinfo {year} {2000})}\BibitemShut {NoStop}%
\bibitem [{\citenamefont {Akazawa}\ \emph {et~al.}(2004)\citenamefont
  {Akazawa}, \citenamefont {Hidaka}, \citenamefont {Fujiwara}, \citenamefont
  {Kobayashi}, \citenamefont {Yamamoto}, \citenamefont {Haga}, \citenamefont
  {Settai},\ and\ \citenamefont {{\=O}nuki}}]{Akazawa2004-ku}%
  \BibitemOpen
  \bibfield  {author} {\bibinfo {author} {\bibfnamefont {T.}~\bibnamefont
  {Akazawa}}, \bibinfo {author} {\bibfnamefont {H.}~\bibnamefont {Hidaka}},
  \bibinfo {author} {\bibfnamefont {T.}~\bibnamefont {Fujiwara}}, \bibinfo
  {author} {\bibfnamefont {T.~C.}\ \bibnamefont {Kobayashi}}, \bibinfo {author}
  {\bibfnamefont {E.}~\bibnamefont {Yamamoto}}, \bibinfo {author}
  {\bibfnamefont {Y.}~\bibnamefont {Haga}}, \bibinfo {author} {\bibfnamefont
  {R.}~\bibnamefont {Settai}}, \ and\ \bibinfo {author} {\bibfnamefont
  {Y.}~\bibnamefont {{\=O}nuki}},\ }\href {\doibase 10.1088/0953-8984/16/4/L02}
  {\bibfield  {journal} {\bibinfo  {journal} {J. Phys. Condens. Matter}\
  }\textbf {\bibinfo {volume} {16}},\ \bibinfo {pages} {L29} (\bibinfo {year}
  {2004})}\BibitemShut {NoStop}%
\bibitem [{\citenamefont {Aoki}\ \emph {et~al.}(2001)\citenamefont {Aoki},
  \citenamefont {Huxley}, \citenamefont {Ressouche}, \citenamefont
  {Braithwaite}, \citenamefont {Flouquet}, \citenamefont {Brison},
  \citenamefont {Lhotel},\ and\ \citenamefont {Paulsen}}]{Aoki2001-vk}%
  \BibitemOpen
  \bibfield  {author} {\bibinfo {author} {\bibfnamefont {D.}~\bibnamefont
  {Aoki}}, \bibinfo {author} {\bibfnamefont {A.}~\bibnamefont {Huxley}},
  \bibinfo {author} {\bibfnamefont {E.}~\bibnamefont {Ressouche}}, \bibinfo
  {author} {\bibfnamefont {D.}~\bibnamefont {Braithwaite}}, \bibinfo {author}
  {\bibfnamefont {J.}~\bibnamefont {Flouquet}}, \bibinfo {author}
  {\bibfnamefont {J.~P.}\ \bibnamefont {Brison}}, \bibinfo {author}
  {\bibfnamefont {E.}~\bibnamefont {Lhotel}}, \ and\ \bibinfo {author}
  {\bibfnamefont {C.}~\bibnamefont {Paulsen}},\ }\href {\doibase
  10.1038/35098048} {\bibfield  {journal} {\bibinfo  {journal} {Nature}\
  }\textbf {\bibinfo {volume} {413}},\ \bibinfo {pages} {613} (\bibinfo {year}
  {2001})}\BibitemShut {NoStop}%
\bibitem [{\citenamefont {Huy}\ \emph {et~al.}(2007)\citenamefont {Huy},
  \citenamefont {Gasparini}, \citenamefont {de~Nijs}, \citenamefont {Huang},
  \citenamefont {Klaasse}, \citenamefont {Gortenmulder}, \citenamefont
  {de~Visser}, \citenamefont {Hamann}, \citenamefont {G{\"o}rlach},\ and\
  \citenamefont {L{\"o}hneysen}}]{Huy2007-th}%
  \BibitemOpen
  \bibfield  {author} {\bibinfo {author} {\bibfnamefont {N.~T.}\ \bibnamefont
  {Huy}}, \bibinfo {author} {\bibfnamefont {A.}~\bibnamefont {Gasparini}},
  \bibinfo {author} {\bibfnamefont {D.~E.}\ \bibnamefont {de~Nijs}}, \bibinfo
  {author} {\bibfnamefont {Y.}~\bibnamefont {Huang}}, \bibinfo {author}
  {\bibfnamefont {J.~C.~P.}\ \bibnamefont {Klaasse}}, \bibinfo {author}
  {\bibfnamefont {T.}~\bibnamefont {Gortenmulder}}, \bibinfo {author}
  {\bibfnamefont {A.}~\bibnamefont {de~Visser}}, \bibinfo {author}
  {\bibfnamefont {A.}~\bibnamefont {Hamann}}, \bibinfo {author} {\bibfnamefont
  {T.}~\bibnamefont {G{\"o}rlach}}, \ and\ \bibinfo {author} {\bibfnamefont
  {H.~V.}\ \bibnamefont {L{\"o}hneysen}},\ }\href {\doibase
  10.1103/PhysRevLett.99.067006} {\bibfield  {journal} {\bibinfo  {journal}
  {Phys. Rev. Lett.}\ }\textbf {\bibinfo {volume} {99}},\ \bibinfo {pages}
  {067006} (\bibinfo {year} {2007})}\BibitemShut {NoStop}%
\bibitem [{\citenamefont {Daido}\ \emph {et~al.}(2019)\citenamefont {Daido},
  \citenamefont {Yoshida},\ and\ \citenamefont {Yanase}}]{daido}%
  \BibitemOpen
  \bibfield  {author} {\bibinfo {author} {\bibfnamefont {A.}~\bibnamefont
  {Daido}}, \bibinfo {author} {\bibfnamefont {T.}~\bibnamefont {Yoshida}}, \
  and\ \bibinfo {author} {\bibfnamefont {Y.}~\bibnamefont {Yanase}},\ }\href
  {\doibase 10.1103/PhysRevLett.122.227001} {\bibfield  {journal} {\bibinfo
  {journal} {Phys. Rev. Lett.}\ }\textbf {\bibinfo {volume} {122}},\ \bibinfo
  {pages} {227001} (\bibinfo {year} {2019})}\BibitemShut {NoStop}%
\bibitem [{\citenamefont {Aroyo}\ \emph {et~al.}(2011)\citenamefont {Aroyo},
  \citenamefont {Perez-Mato}, \citenamefont {Orobengoa}, \citenamefont {Tasci},
  \citenamefont {de~la Flor},\ and\ \citenamefont {Kirov}}]{Aroyo2011-kj}%
  \BibitemOpen
  \bibfield  {author} {\bibinfo {author} {\bibfnamefont {M.~I.}\ \bibnamefont
  {Aroyo}}, \bibinfo {author} {\bibfnamefont {J.~M.}\ \bibnamefont
  {Perez-Mato}}, \bibinfo {author} {\bibfnamefont {D.}~\bibnamefont
  {Orobengoa}}, \bibinfo {author} {\bibfnamefont {E.}~\bibnamefont {Tasci}},
  \bibinfo {author} {\bibfnamefont {G.}~\bibnamefont {de~la Flor}}, \ and\
  \bibinfo {author} {\bibfnamefont {A.}~\bibnamefont {Kirov}},\ }\href@noop {}
  {\bibfield  {journal} {\bibinfo  {journal} {Bulg. Chem. Commun.}\ }\textbf
  {\bibinfo {volume} {43}},\ \bibinfo {pages} {183} (\bibinfo {year}
  {2011})}\BibitemShut {NoStop}%
\bibitem [{\citenamefont {Aroyo}\ \emph
  {et~al.}(2006{\natexlab{a}})\citenamefont {Aroyo}, \citenamefont
  {Perez-Mato}, \citenamefont {Capillas}, \citenamefont {Kroumova},
  \citenamefont {Ivantchev}, \citenamefont {Madariaga}, \citenamefont {Kirov},\
  and\ \citenamefont {Wondratschek}}]{Aroyo2006-bn}%
  \BibitemOpen
  \bibfield  {author} {\bibinfo {author} {\bibfnamefont {M.~I.}\ \bibnamefont
  {Aroyo}}, \bibinfo {author} {\bibfnamefont {J.~M.}\ \bibnamefont
  {Perez-Mato}}, \bibinfo {author} {\bibfnamefont {C.}~\bibnamefont
  {Capillas}}, \bibinfo {author} {\bibfnamefont {E.}~\bibnamefont {Kroumova}},
  \bibinfo {author} {\bibfnamefont {S.}~\bibnamefont {Ivantchev}}, \bibinfo
  {author} {\bibfnamefont {G.}~\bibnamefont {Madariaga}}, \bibinfo {author}
  {\bibfnamefont {A.}~\bibnamefont {Kirov}}, \ and\ \bibinfo {author}
  {\bibfnamefont {H.}~\bibnamefont {Wondratschek}},\ }\href {\doibase
  10.1524/zkri.2006.221.1.15} {\bibfield  {journal} {\bibinfo  {journal} {Z.
  Krist.}\ }\textbf {\bibinfo {volume} {221}},\ \bibinfo {pages} {15} (\bibinfo
  {year} {2006}{\natexlab{a}})}\BibitemShut {NoStop}%
\bibitem [{\citenamefont {Aroyo}\ \emph
  {et~al.}(2006{\natexlab{b}})\citenamefont {Aroyo}, \citenamefont {Kirov},
  \citenamefont {Capillas}, \citenamefont {Perez-Mato},\ and\ \citenamefont
  {Wondratschek}}]{Aroyo2006-jv}%
  \BibitemOpen
  \bibfield  {author} {\bibinfo {author} {\bibfnamefont {M.~I.}\ \bibnamefont
  {Aroyo}}, \bibinfo {author} {\bibfnamefont {A.}~\bibnamefont {Kirov}},
  \bibinfo {author} {\bibfnamefont {C.}~\bibnamefont {Capillas}}, \bibinfo
  {author} {\bibfnamefont {J.~M.}\ \bibnamefont {Perez-Mato}}, \ and\ \bibinfo
  {author} {\bibfnamefont {H.}~\bibnamefont {Wondratschek}},\ }\href {\doibase
  10.1107/S0108767305040286} {\bibfield  {journal} {\bibinfo  {journal} {Acta
  Cryst. A}\ }\textbf {\bibinfo {volume} {62}},\ \bibinfo {pages} {115}
  (\bibinfo {year} {2006}{\natexlab{b}})}\BibitemShut {NoStop}%
\bibitem [{\citenamefont {Elcoro}\ \emph {et~al.}(2021)\citenamefont {Elcoro},
  \citenamefont {Wieder}, \citenamefont {Song}, \citenamefont {Xu},
  \citenamefont {Bradlyn},\ and\ \citenamefont {Bernevig}}]{Elcoro2021-gs}%
  \BibitemOpen
  \bibfield  {author} {\bibinfo {author} {\bibfnamefont {L.}~\bibnamefont
  {Elcoro}}, \bibinfo {author} {\bibfnamefont {B.~J.}\ \bibnamefont {Wieder}},
  \bibinfo {author} {\bibfnamefont {Z.}~\bibnamefont {Song}}, \bibinfo {author}
  {\bibfnamefont {Y.}~\bibnamefont {Xu}}, \bibinfo {author} {\bibfnamefont
  {B.}~\bibnamefont {Bradlyn}}, \ and\ \bibinfo {author} {\bibfnamefont
  {B.~A.}\ \bibnamefont {Bernevig}},\ }\href {\doibase
  10.1038/s41467-021-26241-8} {\bibfield  {journal} {\bibinfo  {journal} {Nat.
  Commun.}\ }\textbf {\bibinfo {volume} {12}},\ \bibinfo {pages} {5965}
  (\bibinfo {year} {2021})}\BibitemShut {NoStop}%
\bibitem [{\citenamefont {Xu}\ \emph {et~al.}(2020)\citenamefont {Xu},
  \citenamefont {Elcoro}, \citenamefont {Song}, \citenamefont {Wieder},
  \citenamefont {Vergniory}, \citenamefont {Regnault}, \citenamefont {Chen},
  \citenamefont {Felser},\ and\ \citenamefont {Bernevig}}]{Xu2020-lo}%
  \BibitemOpen
  \bibfield  {author} {\bibinfo {author} {\bibfnamefont {Y.}~\bibnamefont
  {Xu}}, \bibinfo {author} {\bibfnamefont {L.}~\bibnamefont {Elcoro}}, \bibinfo
  {author} {\bibfnamefont {Z.-D.}\ \bibnamefont {Song}}, \bibinfo {author}
  {\bibfnamefont {B.~J.}\ \bibnamefont {Wieder}}, \bibinfo {author}
  {\bibfnamefont {M.~G.}\ \bibnamefont {Vergniory}}, \bibinfo {author}
  {\bibfnamefont {N.}~\bibnamefont {Regnault}}, \bibinfo {author}
  {\bibfnamefont {Y.}~\bibnamefont {Chen}}, \bibinfo {author} {\bibfnamefont
  {C.}~\bibnamefont {Felser}}, \ and\ \bibinfo {author} {\bibfnamefont {B.~A.}\
  \bibnamefont {Bernevig}},\ }\href {\doibase 10.1038/s41586-020-2837-0}
  {\bibfield  {journal} {\bibinfo  {journal} {Nature}\ }\textbf {\bibinfo
  {volume} {586}},\ \bibinfo {pages} {702} (\bibinfo {year}
  {2020})}\BibitemShut {NoStop}%
\bibitem [{\citenamefont {Sun}\ and\ \citenamefont
  {Linder}(2023)}]{Sun2022-ri}%
  \BibitemOpen
  \bibfield  {author} {\bibinfo {author} {\bibfnamefont {C.}~\bibnamefont
  {Sun}}\ and\ \bibinfo {author} {\bibfnamefont {J.}~\bibnamefont {Linder}},\
  }\href {\doibase 10.1103/PhysRevB.107.144504} {\bibfield  {journal} {\bibinfo
   {journal} {Phys. Rev. B}\ }\textbf {\bibinfo {volume} {107}},\ \bibinfo
  {pages} {144504} (\bibinfo {year} {2023})}\BibitemShut {NoStop}%
\end{thebibliography}%

\end{document}